\documentclass[11pt,a4paper]{article}
\usepackage{jstyle}
\usepackage{multirow}
\usepackage{tikz-cd}
\usetikzlibrary{arrows}

\newcommand{\C}{\mathbb{C}}

\newcommand{\R}{\mathbb{R}}

\newcommand{\rd}{\mathrm{d}}

\author{Euihun JOUNG}
\author{\quad TaeHwan OH}

\affiliation{Department of Physics and Research Institute of Basic
  Science, \\ Kyung Hee University, Seoul 02447, Korea}

\emailAdd{euihun.joung@khu.ac.kr}
\emailAdd{hadron@khu.ac.kr}

\title{\centering
Manifestly Covariant Worldline Actions \\ from Coadjoint Orbits  \\
{\LARGE Part II: Twistorial Descriptions}}

\abstract{
Worldline actions for various twistor particles
in AdS spacetimes are constructed
from the coadjoint orbits
of $Sp(4,\mathbb R)$, $SU(2,2)$
and $O^*(8)$
as constrained Hamiltonian systems.
The constraints are associated with
the coadjoint orbits of the dual groups,
respectively identified as
$O(p,M-p)$, $U(p, M-p)$ and $Sp(p, M-p)$.
These actions are presented in a universal form by making use of the twistor variables 
taking value in $\mathbb R$,
$\mathbb C$ and $\mathbb H$, respectively.
We find that the massless, massive
and the conformal particles (namely the singletons) of any spins appear for the compact dual groups with $p=0$
reproducing many of the results
in earlier literature,
whereas more exotic particle species,
such as various tachyons, continuous spin particles and BdS particles appear 
for the non-compact dual groups.

}

\begin{document}

\maketitle

\section{Introduction}
In this article, we continue the construction 
\cite{Basile:2023vyg} of worldline particle actions 
to the twistor particles.
To begin with, let us summarize briefly the previous work \cite{Basile:2023vyg},
where we classified the coadjoint orbits of the isometry group $G$
of Minkowski, dS and AdS spacetimes,
namely, $ISO(1,d-1)\,,\ SO(2,d-1)\,,$ and $SO(1,d)$,
and interpreted them as different particle species.
This is based on 
Wigner's idea
that fundamental particles carry
unitary irreducible representations
of the isometry group,
and the orbit method
which allows to construct unitary irreducible representations out of coadjoint orbits (see e.g. \cite{Kirillov1999}).
This viewpoint of the relativistic particle dynamics has been proposed already in early literature,
 e.g. \cite{Souriau2012, Balachandran:1983oit,Howe:1992bv}.
In this paper, we use the term \emph{particles}
 to designate coadjoint orbits,
even in the case where the physical property
for particle interpretation is rather unclear,
such as various tachyons, continuous spin particles
as well as the ones living in the bitemporal AdS space, defined as
$\eta_{AB}\,X^A\,X^B=1$ with $\eta={\rm diag}(-1,-1,1,\ldots,1)$, that we abbreviate as BdS space.\footnote{In a sense,
it is also natural to regard it as the bitemporal dS space. See
\cite{Hinterbichler:2023nyz, Pathak:2024cpo}
for the appearance of BdS in other contexts.}

The construction 
of worldline action is given by the pullback of the KKS symplectic potential,
which we reformulate as constrained Hamiltonian systems by implementing
the defining conditions of 
the isometry group as constraints.
Depending on the choice of the 
coadjoint orbits of the isometry group $G$,
we find different constrained systems
where the constraints are associated with
coadjoint orbits of a Lie group $\tilde G$, which we refer to as dual group.  
This choice of name follows from the reductive dual pair correspondence which arises upon quantizing this constrained Hamiltonian system.
We found that the dual group of $G=ISO(1,d-1)$ is $\tilde G=Sp(2M,\mathbb R)
\ltimes Heis_{2M}$,
and the dual group of $G=SO(1,d)$ and $SO(2,d-1)$ is $Sp(2(M+1),\mathbb R)$, for both cases.
Here, the number $M$ is determined by the label of the $G$-coadjoint orbit: for the example of massive spinning particle, it is the number of rows in the
Young diagram representing the spin.

For special dimensions, the 
double covers of the AdS group $SO(2,d-1)$
are isomorphic to other classical Lie groups:
$\widetilde{SO}^+(2,3)\cong Sp(4,\mathbb R)$,
$\widetilde{SO}^+(2,4)\cong SU(2,2)$
and 
$\widetilde{SO}^+(2,6)\cong O^*(8)$.
Therefore, the worldline particle actions of these groups can be constructed
by using different defining conditions,
and hence, different dual groups.
We find that
the dual groups of $G=Sp(4,\mathbb R)$,
$SU(2,2)$ and $O^*(8)$ 
are respectively $\tilde G=O(p,M-p)$,
$U(p,M-p)$ and $Sp(p,M-p)$.
The resulting worldline particle actions
comprise various twistor particles
constructed earlier in the literature,
together with several new twistorial models
for more general types of particles.

Descriptions of worldline particles
using twistors \cite{Penrose:1967wn,Penrose:1972ia}
have long history:
massless/massive 
and scalar/spinning twistor particles in flat/AdS spacetime with/without SUSY
have been addressed in literature.
These worldline models share the same symplectic structures 
\cite{Tod:1977vf},
but feature slightly different constraints.
Anyway, in order to describe a massive particle,
the constraints should depend explicitly on the mass label.
On the other hand, to describe a spinning particle,
often super \emph{scalar} particles are considered where the number of SUSY is
proportional to the maximum spin that the
super particle contains.
In fact, it is also possible to describe a spinning particle by
deforming the constraints
so that they depend explicitly on the spin labels. Our construction follows
the latter method.

The twistor description for worldline particle action was introduced first for the 4d massless (super) conformal scalar particle in \cite{Ferber:1977qx, Shirafuji:1983zd, Bengtsson:1987si},
and the model was 
generalized to 3d and 6d in \cite{Bengtsson:1987ap, Townsend:1991sj, Howe:1992bv}.
Note that the last paper \cite{Howe:1992bv} makes use of the same
coadjoint method that we used in this paper, so our work can be regarded as its generalization to 
general particle species.
See also \cite{Cederwall:1993xe}
for related discussions with more emphasis on the role of division algebra.

After understanding that
a single complex twistor describes 4d massless scalar,
it was subsequently understood that two complex twistors
can describe a massive particle in Mink$_4$ 
\cite{Bette:1984qt}:
the relevant Mink$_{3,4}$ 
twistor models
with explicit mass and spin label
are studied in \cite{Fedoruk:2003td,
Bette:2005br,Fedoruk:2005ks} (see also more recent works 
 in this direction, \cite{Mezincescu:2013nta}
and \cite{Fedoruk:2014vqa},
for super scalar and spin $s$ particles, respectively)
where the constraints are quadratic in twistor variables and 
involve the mass and spin label as constant shifts.
This is the feature that our particle models share.
Different two-twistor massive particle models in Mink$_3$, Mink$_6$ and Mink$_{3,4,6}$ are considered 
respectively in \cite{Mezincescu:2010yq}, \cite{Routh:2015ifa} and \cite{Mezincescu:2015apa}.
In these models,
the mass constraint is given by
a determinant condition so they are quartic 
in twistor variables,
and non-zero spins are included by SUSY instead of deforming the constraints by spin labels.

The discovery of AdS/CFT
incited  motivation  to 
explore
the twistor description for particles 
living in AdS spacetime. 
In \cite{Claus:1999zh,Zunger:2000wy}, a complex two-twistor model has been shown to describe massive scalar particles in AdS$_5$.
This model is in fact rather exceptional because 
in general massive AdS particles require four twistors instead of two, as shown for  AdS$_{4,5,7}$ massive scalar particles in \cite{Cederwall:2000km},
and consistent with
the general pattern found in the analysis of relevant oscillator representations   \cite{Claus:1999xr}.
See also \cite{Uvarov:2016slb,
Uvarov:2018ose,Uvarov:2019vmd}
for the AdS$_{5}\times S^5$ massless super twistor particles.
Two-twistor models for the massless AdS$_{4,5,7}$ super particle 
are studied
in \cite{Arvanitakis:2016wdn,
Arvanitakis:2016vnp,
Arvanitakis:2017cpk} in relation with the Mink$_{3,4,6}$ massive twistors 
\cite{Mezincescu:2010yq,Routh:2015ifa,Mezincescu:2015apa},
where the AdS$_5$ exception is  clarified further.
See also  \cite{Koning:2023ruq,Koning:2024vdl} 
for another twistorial (Lagrangian) description of the AdS massive particles.

As we explained before, in order to describe a definite spin $s$ particle,
a ``spin-shell'' constraint must be imposed. If we remove this constraint in a proper manner, the action can describe infinitely many particles
of any allowed spins, and
hence provides a worldline model 
for higher spin theory. 
Such models for massless higher spin particles have been considered in
\cite{Cederwall:2004cf, Fedoruk:2005np,Fedoruk:2006it,Fedoruk:2007ud} and \cite{Fedoruk:2013sna},
and massive higher spin models 
are also studied in \cite{deAzcarraga:2014hda}
and \cite{Kim:2021rda}.
See also \cite{Townsend:1991yq,Uvarov:2014lfa}
for the use of classical twistor variables in constructing higher spin algebras.

Twistor descriptions are not limited to
the ordinary massless and massive (spinning) particles,
but they can cover also the continuous spin particles: the Mink$_4$ 
and Mink$_6$
continuous spin particle actions are 
obtained respectively in  \cite{Buchbinder:2018soq,Buchbinder:2019iwi,Buchbinder:2019sie}
and \cite{Buchbinder:2021bgv}.

\medskip

The current paper is organized as follows.
In Section \ref{sec:classical action}, 
we begin with the review of the isomorphisms
between $\widetilde{SO}^+(2,d)$
and ``twistor groups'':
$Sp(4,\mathbb R)$, $SU(2,2)$
and $O^*(8)$ for $d=3,4$ and $6$.
Then we demonstrate how 
various twistor particle actions
can be obtained from coadjoint orbits 
of twistor groups,
by implementing their defining condition as 
Hamiltonian constraints.
The constraints are associated 
with the coadjoint orbits of dual groups,
respectively identified with
$O(p,M-p)$, $U(p,M-p)$ and $Sp(p,M-p)$,
for $d=3,4$ and 6,
where $M$ is the number of twistors.
In Section \ref{sec:ads action},
for each of $\mathfrak{so}(2,d)$ coadjoint orbits 
in the list of classification 
obtained in \cite{Basile:2023vyg},\footnote{See \cite{Stern:2009jh} for
the 4d orbit classification 
based on $GL(2,\mathbb C)$.}
we determine the system of constraints,
associated with the dual groups, of the corresponding twistor actions.
We find dual groups with different $M$ and $p$ appear
for different coadjoint orbits, namely for different particle species.
In Section \ref{sec:summary},
we summarize the results according to the dual groups,
and generalize them by making use of the 
dual pair correspondence \cite{Basile:2023vyg}.
For compact dual groups with $M=1$, 
we find conformal particles in $|\mathbb F|+2$ dimensions,
and the $M=2$ and $4$ cases describe massless and massive particles in AdS$_{|\mathbb F|+3}$, respectively,
while
the $M=3$ case corresponds to ``very light'' massive particles.
Non-compact dual groups cover all other exotic particle
species such as continuous spin particles, tachyons, BdS particles etc.
In particular, the continuous spin particles
arise for the $M=4$ and $p=1$ case. 
In Section \ref{sec:restriction},
we show how the Mink$_{|\mathbb F|+2}$ or (A)dS$_{|\mathbb F|+2}$ massive twistor particle actions
can be obtained by 
restricting the massless one of (A)dS$_{|\mathbb F|+3}$.
This discussion is along the line of \cite{Arvanitakis:2017cpk}, while we make more emphasizes
on the algebraic side, such as see-saw dual pairs.
Finally, in Appendix \ref{sec:quaternion},
we provides complex matrix representations 
of quaternion and their use in Poisson algebra,
and Appendix \ref{sec:ads dic}
contains the explicit component expressions of 
the AdS generators,
which can serve as a dictionary between
vectorial and twistorial descriptions.

\section{Manifestly covariant worldline actions 
from coadjoint orbits of a classical Lie group}
\label{sec:classical action}
Let us review the construction of manifestly covariant particle actions 
from coadjoint orbits of a classical Lie group \cite{Basile:2023vyg}.

\subsection{Classical Lie Groups}

Any classical Lie group is a  subgroup of a general linear group 
$GL(N, \mathbb F)$ with $\mathbb F=\mathbb R, \mathbb C$ and $\mathbb H$,
which preserves a certain bilinear form:
We can define any classical Lie group  as
\begin{equation}
    B(b_{\sst (N)},\mathbb F)
   = \{\,A \in GL(N,\mathbb F) \mid 
    A^\dagger\,b_{\sst (N)}\,A = b_{\sst (N)}\,\}\,,
    \label{B group}
\end{equation}
where $b_{\sst (N)}$ is an element of $GL(N,\mathbb F)$\,.
The dagger operation is defined as the composition $A^\dagger=(A^{t})^*$ 
of the matrix transpose $t$ and the $\mathbb F$  conjugation $*$.%
\footnote{The $*$ conjugation is the identity map
for $\mathbb R$, the complex conjugation for $\mathbb C$,
and the quaternionic conjugation for $\mathbb H$\,.}
Up to a $GL(N,\mathbb F)$ conjugation,
$b_{\sst (N)}$ can be mapped to either Hermitian or anti-Hermitian 
with respect to $\dagger$. A Hermitian $b_{\sst (N)}$ is in the conjugacy class of a flat metric of $(p,N-p)$ signature: $b_{\sst (N)}=\eta_{\sst (p,N-p)}$.
An anti-Hermitian $b_{\sst (N)}$ is 
in the conjugacy class of  the symplectic matrix: 
$b_{\sst (N)}=\Omega_{\sst (N)}$\,. With different choices 
of $b_{\sst (N)}$, we find the full list of classical Lie groups,
besides the $GL(N,\mathbb F)$ itself:
\begin{eqnarray}
    & B(\eta_{\sst (p,N-p)},\mathbb R) = O(p,N-p)\,,
    \qquad
    & B(\Omega_{\sst (N)},\mathbb R) = Sp(N,\mathbb R)\,,\\
	& B(\eta_{\sst (p,N-p)},\mathbb C) = U(p,N-p)\,,
    \qquad 
    & B(\Omega_{\sst (N)},\mathbb C) \cong U(\tfrac{N}2,\tfrac{N}2)\,,\\
    & B(\eta_{\sst (p,N-p)},\mathbb H) = Sp(p,N-p)\,,
    \qquad 
    & B(\O_{\sst (N)},\mathbb H) \cong O^*(2N)\,.
\end{eqnarray}
In the right column, $N$ is even for $\mathbb F=\mathbb R$ and $\mathbb C$,
whereas it can be also an odd number for $\mathbb H$.
This is because we can take  the second element of $\mathbb H$ 
as a two-dimensional symplectic matrix:
$\O_{\sst (N)}=\mathsf{j}\,I_{\sst (N)}$.
Since $\O_{\sst (N)}$ can be diagonalized as $\mathsf{i}\,\eta_{\sst (N/2,N/2)}$,
the group
$B(\O_{\sst (N)},\mathbb C)$ is isomorphic to $U(N/2,N/2)$.
On the contrary,  even though $\O_{\sst (N)}$ can still be diagonalized 
as $\mathsf{i}\,\eta_{\sst (N/2,N/2)}$, the group $B(\O_{\sst(N)},\mathbb H)$ 
is not isomorphic to $Sp(N/2,N/2)$ because the $\mathsf{i}$, 
the first element of $\mathbb H$, does not commute with a quaternionic matrix.
See e.g. \cite[Prop. 9.3.2.]{Collingwood1993} for relevant discussions.

\subsection{Twistor Groups}

{The classical Lie groups $B(\O_{\sst (4)},\mathbb F)$
 happen to be isomorphic to
AdS groups of special dimensions,
\begin{eqnarray}
	&& B(\O_{\sst (4)},\R)=Sp(4,\R)\cong \widetilde{SO}^+(2,3)\,,\nn 
	&& B(\O_{\sst (4)},\C)=U(2,2) \supset SU(2,2)\cong \widetilde{SO}^+(2,4)\,,\nn
	&& B(\O_{\sst (4)},\mathbb H)=O^*(8)\cong \widetilde{SO}^+(2,6)\,,
	\label{AdS isomorphism}
\end{eqnarray}
and they 
contain $GL(2,\mathbb F)$
isomorphic 
with the Lorentz subgroups,
\begin{eqnarray}
	&& GL(2,\R)\supset SL(2,\R)\cong  \widetilde{SO}^+\!(1,2)\,,\nn 
	&& GL(2,\C)\supset SL(2,\C) \cong \widetilde{SO}^+\!(1,3)\,,\nn
	&& GL(2,\mathbb H)\supset SL(2,\mathbb H) \cong \widetilde{SO}^+\!(1,5)\,.
	\label{dS isomorphism2}
\end{eqnarray}
Here $\widetilde{SO}^+(p,q)$ designates the double cover 
of the orthochronous Lorentz group $SO^+(p,q)$.
Hence,
letting $|\mathbb F|$ be
the real dimensions of $\mathbb F$,
the twistor group
$B(\O_{\sst(4)},\mathbb F)$
is the double cover of the AdS$_{|\mathbb F|+2}$ isometry group, or equivalently the $(|\mathbb F|+1)$-dimensional 
conformal group. 
This isomorphism allows to describe half-integral spins,
which was not possible in the vectorial description 
based on the Lie group $O(p,q)$ \cite{Basile:2023vyg},
and it is in the heart of twistor description of particles 
and fields: see e.g. \cite{Howe:1992bv, Cederwall:2000km}.
We shall show how the twistor actions of various AdS particles 
in the literature can be recovered, together with many new 
particle actions,
from a universal expression  by applying 
the information of Lie group, that is $b_{\sst (N)}$, 
and the coadjoint orbit $\cO^G_\phi$, which
will be specified by a representative element $\phi$. 

All the twistor actions will be described 
in an equal footing by taking the advantage of the natural inclusion
as groups of $4 \times 4$ matrices,
\begin{equation}
    B(\O_{\sst(4)},\mathbb R)
    \subset
    B(\O_{\sst(4)},\mathbb C)
    \subset 
    B(\O_{\sst(4)},\mathbb H)\,,
    \label{inclusion}
\end{equation}
with the $2 \times 2$ matrix Lorentz subgroups,
\begin{equation}
	GL(2,\R) \subset GL(2,\C) \subset GL(2,\mathbb H)\,.
\end{equation}
This allows us to describe the corresponding twistor actions 
in a universal expression in terms of $4M\,|\mathbb F|$ variables,
where $M$ is the rank of the dual group fixed by
the choice of the coadjoint orbit $\cO^G_\phi$.


\subsection{$GL(N,\mathbb F)$ manifestly covariant worldline action}

Let us now review  the derivation of a manifestly covariant worldline action 
with a $GL(N,\mathbb F)$ symmetry 
from a coadjoint orbit of the Lie group $GL(N,\mathbb F)$.
On the group manifold $GL(N,\mathbb F)$, the matrix components
of $X^a{}_{b}\in \mathbb F$ of an element $X \in  GL(N,\mathbb F)$
serve a natural and global coordinate system,
and the left-invariant vector fields are given by
$V=V^a{}_b\,X^{c}{}_a\,\frac{\partial}{\partial X^c{}_b}$.
The symplectic structure of the cotangent bundle $T^*GL(N,\mathbb F)$
prescribes the geometric action, which reads simply
\begin{equation}
    S[X,P,A] = \int\tr_{N}\left[P\,\rd X
    + A\,(P\,X - \phi) \right]_{\mathbb R},
\end{equation}
where  $X$, $P$, $A$ and $\phi$ belong to $Mat_{N\times N}(\mathbb F)$.
The real part $[-]_{\mathbb R}$ stands for 
\be
	\left[A\right]_{\mathbb R}=
	\frac12\left(A+A^\dagger\right),
\ee
with $A^\dagger:=(A^t)^*$, and the $(-)^*$ stands for the conjugate in $\mathbb F$. 
The real part does not do anything for $\mathbb F=\mathbb R$, but they are necessary 
for the correct symplectic potentials for $\mathbb F = \mathbb C$ 
and $\mathbb F = \mathbb H$. The constraint imposed 
by the Lagrange multiplier $A$ can be algebraically solved to give
\begin{equation}
	S[X] = \int \tr_{N} \left[\phi\,X^{-1} \rd X\right]_{\mathbb R}\,.
    \label{mat g inv}
\end{equation}
If the coadjoint matrix $\phi$ has a rank $M\le N$,
we can transform it into a triangular form
and integrate out the components of $X$ in the lower subspace
$Mat_{N \times (N-M)}(\mathbb F)$.
This brings the reduction of  the action into the form,
 \be
	S[X,P,A] = \int\tr_{M}\left[P\,\rd X
    + A\,(P\,X - \tilde\phi) \right]_{\mathbb R}\,,
\ee 
where $X$, $P$ and $A$ takes value 
in $Mat_{N\times M}(\mathbb F)$,
$Mat_{M\times N}(\mathbb F)$
and $Mat_{M\times M}(\mathbb F)$, respectively.
The matrix $\tilde \phi$ is the $M\times M$
submatrix of the triangulized $\phi$\,,
and hence it belongs to $Mat_{M\times M}(\mathbb F)$\,.
In the end, the action describes a $G=GL(N,\mathbb F)$
coadjoint orbit $\cO^G_\phi$ as a reduced phase space
inside $\mathbb F^{2MN}$ where the constraints
are given through the moment maps
$\tilde\mu(X,P)=P\,X\in Mat_{M\times M}(\mathbb F)$
generating $\tilde G=GL(M,\mathbb F)$ under Poisson bracket.
Note also that the moment maps
$\mu(X,P)=X\,P\in Mat_{N\times N}(\mathbb F)$
associated with the original $GL(N,\mathbb F)$ symmetry
Poisson-commute with $\tilde\mu$\,. 
The constraints $\tilde\mu(X,P) - \tilde\phi \approx 0$
are associated with the dual $\tilde G=GL(M,\mathbb F)$
coadjoint orbit $\cO^{\tilde G}_{\tilde\phi}$.

\subsection{$B(b_{\sst (N)},\mathbb F)$ manifestly 
covariant worldline action}

The passage from $GL(N,\mathbb F)$ to $B(b_{\sst (N)},\mathbb F)$
can be done by implementing the defining condition \eqref{B group} 
of $B(b_{\sst (N)},\mathbb F)$ into the action \eqref{mat g inv} 
as a constraint. The resulting action reads
\begin{eqnarray}
	S[X,A] \eq \int \tr_{N}\left[\phi\,X^{-1} \rd X
    + A\,(X^\dagger\,b_{\sst (N)}\,X-b_{\sst(N)})\right]_{\mathbb R}\nn
    & \cong & \int \tr_{N}\left[\phi\,
    b_{\sst (N)}^{\ \ -1}\,X^\dagger\,b_{\sst (N)}\,\rd X
    + A\,(X^\dagger\,b_{\sst (N)}\,X - b_{\sst (N)})\right]_{\mathbb R}\,,
\end{eqnarray}
where $\cong$ designates the redefinition of $A$ 
replacing $X^{-1}$ by $b_{\sst (N)}^{\ \ -1}\,X^\dagger\,b_{\sst (N)}$. 
The constant matrix $\phi$ belongs to
\begin{equation}
	\mathfrak{b}(b_{\sst (N)},\mathbb F)
	=\{\,\xi\in Mat_{N\times N}(\mathbb F)\,|\,
	\xi^\dagger\,b_{\sst (N)}+b_{\sst(N)}\,\xi=0\,\}\,,
\end{equation}
whereas $X$ takes value in $Mat_{N\times N}(\mathbb F)$ 
after introducing the constraint. 
If the matrix $\phi\in \mathfrak{b}(b_{\sst (N)},\mathbb F)$ 
has a rank $M$, it can always be mapped to a $\tilde b_{\sst (M)}$ as
\begin{equation}
    \phi = T\,\tilde b_{\sst (M)}\,T^\dagger\,b_{\sst (N)}\,,
\end{equation}
in terms of a transformation matrix $T\in Mat_{N\times M}(\mathbb F)$.
Moreover, the mapping has a very simple rule:
we can always choose $T$ such that
$\tilde b_{\sst (M)}=(\O_{\sst (M)})^{-1}$ 
for $b_{\sst (N)}=\eta_{\sst (p,N-p)}$,
and $\tilde b_{\sst (M)}=(\eta_{\sst (q,M-q)})^{-1}$
for  $b_{\sst (N)}=\O_{\sst (N)}$. 
In the latter case, the number $q$ depends on $\phi$.
The real part
concerns mostly about the quaternion case 
since in the real and complex cases
the Lagrangian is already real up to a boundary term.

After a suitable redefinition of $X$ and $A$ 
in terms of the transformation matrix $T$
and integrating out the non-dynamical components of $X$
in the subspace $Mat_{N\times (N-M)}(\mathbb F)$,
the action becomes
\begin{equation}
	S[X,A] = \int \tr_{M}\left[\tilde b_{\sst (M)}\,
	X^\dagger\,b_{\sst (N)}\,\rd X +A\left(X^\dagger\,b_{\sst (N)}\,X
	-\tilde b_{\sst (M)}^{\ \ -1}\,\tilde\phi\right)\right]_{\mathbb R}\,,
	\label{universal}
\end{equation}
where $X$ takes value in $Mat_{N\times M}(\mathbb F)$ 
and $\tilde\phi$ is given by
\begin{equation} \label{dual orbit}
   \tilde b_{\sst (M)}^{\ \ -1}\,\tilde\phi =T^\dagger\,b_{\sst (N)}\,T\,.
\end{equation}
Here, $\tilde \mu(X)=X^\dagger\,b_{\sst (N)}\,X$
are the moment maps  
generating the dual symmetry $\tilde G=B(\tilde b_{\sst (M)},\mathbb F)$,
and hence $\tilde\phi$ is an element 
of $\mathfrak{b}(\tilde b_{\sst (M)},\mathbb F)$\,.
Note that the moment maps $\tilde \mu$ of the dual group Poisson-commute with the moment maps
$\mu(X) = X\,\tilde b_{\sst (M)}\,X^\dagger$
generating the original symmetry $G=B(b_{\sst (N)},\mathbb F)$\,.

\subsection{Worldline action
of twistor particles in AdS$_{|\mathbb F|+2}$}

With the choice $b_{\sst (4)}=\O_{\sst (4)}$
and $(\tilde b_{\sst (M)})^{-1}=\eta_{\sst (q,M-q)}$,
the universal action \eqref{universal} can be written 
in a component form as
\begin{equation}
	S[X,A] = \int \left[ \eta^{IJ}\,
	(X^A{}_J)^*\,\Omega_{AB}\,\rd X^B{}_I
	+A^{IJ}\,\left((X^A{}_J)^*\,\Omega_{AB}\,X^B{}_I
	-\eta_{JK}\,\tilde\phi^{K}{}_{I}\right) \right]_{\mathbb R},
\end{equation}
where  $\eta_{IJ}$ are the components of $\eta_{\sst (q,M-q)}$.
The dual group indices $I, J$ take values in
$\{1, \ldots, M\}$ for $\eta_{\sst (M)}$,
$\{0, 1, \ldots, M-1\}$ for $\eta_{\sst (1,M-1)}$
and 
$\{0',0,1\ldots, M-2\}$ for $\eta_{\sst (2,M-2)}$.
These indices should not be confused with the spacetime coordinate indices.
What transforms covariantly under the spacetime symmetry
is the twistor variable with the indices $A, B$,
which takes 4 values. They can be further split into the undotted 
and dotted bi-valued indices $\a,\b$ and $\dot\a, \dot\b$:
\begin{equation}
	\Omega_{AB} = \begin{pmatrix} 
	0 & -\delta^{\dot\beta}_{\dot\a} \\
	\delta^{\alpha}_{\beta} & 0
	\end{pmatrix}\,,
	\qquad X^A{}_I=\binom{\xi^\alpha{}_I}{\pi_{\dot\alpha}{}_I}\,,
	\qquad (X^A{}_I)^*=\binom{(\xi^\alpha{}_I)^*}{(\pi_{\dot\alpha}{}_I)^*}
	=\binom{\bar\xi^{\dot\alpha}{}_I}{\bar\pi_{\alpha}{}_I}\,,
\end{equation}
to result in the more familiar form of the twistor action,
\begin{equation}
	S[\xi,\pi,A] = \int \left[\eta^{IJ}\,
	\Big(\bar \pi_{\a\,J}\,\rd \xi^\alpha{}_I
	-\bar\xi^{\dot\a}{}_J\,\rd\pi_{\dot\a\,I}\Big)
	+A^{IJ}\,\left(\bar\pi_{\a\,J}\, \xi^\alpha{}_I
	-\bar\xi^{\dot\a}{}_J\,\pi_{\dot\a\,I}
	-\tilde\phi_{JI}\right)\right]_\R.
	\label{AdS twistor action}
\end{equation}
Here, $(\tilde \phi_{IJ})^*=-\tilde\phi_{JI}$ 
and $(A^{IJ})^*=-A^{JI}$ under the $\mathbb F$ conjugation.
For $\mathbb F=\mathbb R$ and $\mathbb C$, the Lagrangian 
without the projection to the real part $[-]_\R$ is already real  up to a boundary term, 
but for $\mathbb F=\mathbb H$, the (conj) term is necessary 
for the reality of the Lagrangian.
Remark that $M$
corresponds to the number of twistors.

The Poisson bracket of the action for $\mathbb F=\mathbb C$ reads
\begin{equation}
	\{\xi^\alpha{}_I, \bar\pi_{\beta\,J}\} 
	= \eta_{IJ}\,\delta_\beta^\alpha\,,
	\qquad 
	\{\bar\xi^{\dot\alpha}{}_I, \pi_{\dot\beta\,J}\}
	=\eta_{IJ}\,\delta_{\dot\beta}^{\dot\alpha}\,,
\end{equation}
where the complex variables and their complex conjugate 
are treated independently as usual. 
For $\mathbb F=\mathbb R$, we may either begin 
with the real variables with
\begin{equation}
	\bar \xi^{\dot\alpha}{}_I=\xi^\alpha{}_I\,,
	\qquad 
	\bar \pi_{\alpha\,I}=\pi_{\dot\alpha\,I}\,,
\end{equation}
or work with the complex variables and treat 
the above reality conditions as second class constraints.
For $\mathbb F=\mathbb H$, 
we may use
the $2\times2$ Hermitian matrix representation of a quaternion, where
the Hermicity condition should be
imposed as second class constraints
(see Section \ref{sec:quaternion}).
In fact, for a universal treatment
of any $\mathbb F$,
it is convenient to make use of the following properties of the real part $[-]_\R$.
For any numbers $a, b\in \mathbb F$, 
\begin{equation}
	[a \,b]_{\mathbb R} = [b\,a]_\R =[a^*\,b^*]_\R
	=a_0\,b_0-\sum_{r=1}^{|\mathbb F|-1} a_r\,b_r\in \mathbb R\,,
\end{equation}
where  the real components $a_r$ are given by
\begin{equation}
	a=a_0\in \mathbb R\,,
	\qquad 
	a=a_0+a_1\,{\mathsf i} \in \mathbb C\,,
	\qquad 
	a=a_0+a_1\,{\mathsf i}+a_2\,{\mathsf j}+a_3\,{\mathsf k}\in \mathbb H\,.
\end{equation}
Then, the Poisson bracket can be expressed  as
\begin{equation}
	\big\{\,[a \, \xi^\alpha{}_I]_\R\,,\, [b \, \bar\pi_{\beta\,J}]_\R\,\big\}
	=\big\{\,[a \, \xi^\alpha{}_I]_\R\,,\, [b^* \, \pi_{\dot\beta\,J}]_\R\,\big\}
	=\frac12\,\eta_{IJ}\,\delta^\alpha_\beta\,[a\, b]_{\R}\,,
\end{equation}
for any $\mathbb F$ numbers  $a$ and $b$. 
The constraints imposed by the Lagrange multiplier $A^{IJ}$ are
\begin{equation}
	\tilde\chi_{IJ}=\tilde\mu_{IJ}-\tilde\phi_{IJ}\approx 0\,, 
\end{equation}
where $\tilde\mu_{IJ}$ are the $\mathbb F$-valued functions given by
\begin{equation} \label{dual gen}
	\tilde\mu_{IJ}=\bar\pi_{\alpha\,I}\, 
	\xi^\alpha{}_J-\bar\xi^{\dot\alpha}{}_I\,\pi_{\dot\alpha\,J}\,,
\end{equation}
and they are closed under the Poisson bracket,
\ba	
	&& \big\{\, [a\,\tilde\mu_{IJ}]_\R\,,\, [b\, \tilde\mu_{KL}]_\R\,\big\}\nn
	&&=\frac12\,\left[\eta_{JK}\,(b\,a)\, \tilde\mu_{IL}
	-\eta_{JL}\,(b^*\,a)\,\tilde\mu_{IK}
	-\eta_{KI}\,(b\,a^*)\,\tilde\mu_{JL}
	+\eta_{LI}\,(b^*\,a^*)\,\tilde\mu_{JK}\right]_\R\,.
\ea
These Poisson brackets 
define the dual algebra $\tilde{\mathfrak{g}}
=\mathfrak{b}(\eta_{\sst (q,M-q)},\mathbb F)$: the functions
$\tilde\mu^{IJ}$ 
are the moment maps of the dual algebra.

Note that for $\mathbb F=\mathbb R$, the real part does not play any role and the
numbers $a, b$ can be factored out freely leaving
the usual Lie bracket of the orthogonal algebra $\mathfrak{so}(q,M-q)$\,.
For $\mathbb F=\mathbb C$, the above is equivalent to the more usual expression,
\be
	\{\tilde\mu_{IJ}\,, \tilde\mu_{KL}\}
	=\eta_{JK}\,\tilde\mu_{IL}-\eta_{IL}\,\tilde\mu_{KJ}\,.
\ee
The moment maps $\tilde\mu_{IJ}$ transform under the $\mathbb F$ conjugate as
\begin{equation}
	(\tilde\mu_{IJ})^*=-\tilde\mu_{JI}\,,
\end{equation}
and hence the symmetric and antisymmetric parts 
$\tilde\mu_{(IJ)}$ and $\tilde \mu_{[IJ]}$ have the property,
\be
	(\tilde\mu_{(IJ)})^*=-\tilde\mu_{(IJ)}\,,
	\qquad 
	(\tilde\mu_{[IJ]})^*=\tilde\mu_{[IJ]}\,.
\ee
This shows that 
each symmetric components $\tilde \mu_{(IJ)}$ are respectively
zero, pure imaginary and pure quaternion for $\mathbb F=\mathbb R, \mathbb C$ and $\mathbb H$.
Therefore, each $\tilde \mu_{(IJ)}$ generates the Lie algebra $\hat{\mathbb F}$
that we define as $\hat{\mathbb R}:=\mathfrak{o}(1)=0$,
$\hat{\mathbb C}:=\mathfrak{u}(1)$ and $\hat{\mathbb H}:=\mathfrak{sp}(1)$ (in other words, $\hat{\mathbb F}=\mathbb F/\R$).

The moment maps  $\tilde\mu_{IJ}$ are 
the image of the co-moment map $m$
of the dual algebra $\tilde{\mathfrak{g}}
=\mathfrak{b}(\eta_{\sst (q,M-q)},\mathbb F)$,
\begin{equation}
	m\left[\frac12\left(a^*\,R_{IJ}-a\,R_{JI}\right)\right]=[a\, \tilde\mu_{IJ}]_\R\,,
	\qquad \forall a \in \mathbb F\,,
\end{equation}
where the basis of $\tilde{\mathfrak{g}}$ is the subset of $\mathbb F\,R_{IJ}$\footnote{By $\mathbb F\,R_{IJ}$,
we mean the basis  $\{R_{IJ}\}$, $\{R_{IJ}, i\,R_{IJ}\}$ 
and $\{R_{IJ}, {\bf i}\,R_{IJ}, {\bf j}\,R_{IJ}, {\bf k}\,R_{IJ}\}$
for  $\mathbb F=\mathbb R$, $\mathbb C$ and $\mathbb H$.}
whose Lie brackets are inherited from the associative product,
\begin{equation}
	R_{IJ}\,R_{KL}=\eta_{JK}\,R_{IL}\,.
\end{equation}
The basis of the coadjoint space $\tilde{\mathfrak{g}}^*=\mathfrak{b}(\eta_{\sst (q,M-q)},\mathbb F)^*$ 
can be chosen in $\mathbb F\,\cR^{IJ}$ 
basis with\footnote{It is convenient to use the defining representations of  
$R_{IJ}$ and $\cR^{IJ}$\,:
\be
      (R_{IJ})^K{}_L=\delta^K_I\,\eta_{JL}\,,
	\qquad
	(\cR^{IJ})^K{}_L=\eta^{IK}\,\delta^J_L\,,
\ee
so that the two are related by $\cR^{IJ}=R^{IJ}=\eta^{IK}\,\eta^{JL}\,R_{KL}$
and satisfy
\be
	\la \cR^{IJ}\,, R_{KL}\ra=
	\tr\left((\cR^{IJ})^t\,R_{KL}\right)=
	\delta^I_K\,\delta^J_L\,.
\ee
}
\be
	\la a\,\cR^{IJ}\,,\,b\,R_{KL}\ra
	=[a^*\,b]_\R\,\delta^I_{K}\,\delta^J_L\,.
\ee
Hence, any elements  $\xi\in\tilde{\mathfrak{g}}$ 
and  $\tilde\phi\in\tilde{\mathfrak{g}}^*$ can be decomposed in the above basis as
\be
	\xi=\xi^{IJ}\,R_{IJ}\,,\qquad \tilde\phi=\tilde\phi_{IJ}\,\cR^{IJ}\,,
\ee 
with $\xi^{IJ}$ and  $\tilde\phi_{IJ}$ in $\mathbb F$ satisfying
\be
	(\xi^{IJ})^*=-\xi^{JI}\,,
	\qquad 
	(\tilde\phi_{IJ})^*=-\tilde\phi_{JI}\,.
\ee
The bilinear form between $\tilde\phi$ and $\xi$ can be expressed as
\be
	\la \,\tilde\phi\,,\,\xi\,\ra=\left[(\tilde\phi_{IJ})^{*}\, \xi^{IJ}\right]_\R= \tr_M\left[ \tilde\f^\dagger\ \x \right]_\R\,.
\ee
The constraints $\chi_{IJ}=\tilde\mu_{IJ}-\tilde\phi_{IJ}\approx 0$
are split into the first and second class constraints.
The first class constraints are the combinations $[\xi^{IJ}\,\chi_{IJ}]_\R$ with
\be
	{\rm ad}^*_{[\xi^{IJ}\,R_{IJ}]_\R}\,\tilde\phi=0\,,
\ee
while the rest corresponding to $\tilde G/{\tilde G}_{\tilde\phi}\cong \cO_{\tilde\phi}^{\tilde G}$ are the second class constraints.
The dimension of the phase space, that is, the original coadjoint orbit $\cO_\phi^G$
is 
\ba
	{\rm dim}(\cO_\phi^G)\eq 4\,M\,|\mathbb F|-
	{\rm dim}(\tilde{\mathfrak{g}})-{\rm dim}(\tilde{\mathfrak{g}}_{\tilde\phi})\nn
	\eq M+\frac{M(7-M)}2\,|\mathbb F|-{\rm dim}(\tilde{\mathfrak{g}}_{\tilde\phi})\,,
 \label{dim orbit}
\ea
where the dimension of $\tilde{\mathfrak{g}}$ is
\be
	{\rm dim}(\tilde{\mathfrak{g}})=\frac{M(M+1)}2\,|\mathbb F|-M\,.
\ee
As we shall see later, the possible values of $M$ are $1, 2, 3, 4$.

Now let us consider the spacetime symmetry which is generated by  the moment maps,
\be
	L^{\a}{}_{\b}=\xi^{\a\,I}\,\bar\pi_{\b\,I}\,,
	\quad P^{\a\dot\b}=\xi^{\a\,I}\,\bar\xi^{\dot\b}{}_{I}\,,
	\quad K_{\dot\a\b}=\pi_{\dot\a}{}^{I}\,\bar\pi_{\b\,I}\,,
	\quad \bar L_{\dot\a}{}^{\dot\b}=\pi_{\dot\a}{}^{I}\,\bar\xi^{\dot\b}{}_I\,,
\ee 
with the Poisson brackets,

\be
    \{\,[a\,L^\a{}]_\b]_\R\,,\,[b\,L^{\g}{}_\d]_\R\,\}
    =\frac{1}{2}\,\left[\delta^\a_\d\,(a\,b)\,L^{\g}{}_\b
	-\delta^\g_\b\,(b\,a)\,L^{\a}{}_\d\right]_\R,
\ee
\be
	\{\,[a\,P^{\a\dot\b}]_\R\,,\,[b\,K_{\dot\g\d}]_\R\,\}
	=\frac{1}{2}\,\left[\delta^\a_\d\,(a\,b)\,\bar L_{\dot\g}{}^{\dot\b}
	+\delta^{\dot\b}_{\dot\g}\,(b\,a)\,L^{\a}{}_\d\right]_\R\,,
\ee
\be
	\{\,[a\,\bar L_{\dot\a}{}^{\dot\b}]_\R\,,\,[b\,\bar L_{\dot\g}{}^{\dot\d}]_\R\,\}
	=\frac{1}{2}\,\left[\delta^{\dot\b}_{\dot\g}\,(b\,a)\,\bar L_{\dot\a}{}^{\dot\d}
	-\delta_{\dot\a}^{\dot\d}\,(a\,b)\,\bar L_{\dot\g}{}^{\dot\b}\right]_\R\,.
\ee
They are the co-moment map image of the Lie algebra $\mathfrak{g}=\mathfrak{b}(\O_{\sst (4)},\mathbb F)$ generated by 
the generators of the conformal algebra.

\section{Classification of  twistor particles in AdS$_{|\mathbb F|+3}$}
\label{sec:ads action}

In this section, we determine the coadjoint matrices $\phi$ corresponding 
to different particle species. For each of $\phi$, we identify $(\tilde b_{\sst (M)})^{-1}=\eta_{\sst (q,M-q)}$,
that defines the dual group, and the dual coadjoint matrix $\tilde\phi$\,.
The classification of coadjoint orbits has been already carried out in the $\mathfrak{so}(2,d)$ basis. 
Therefore, we simply need to transform the  $\mathfrak{so}(2,d)$ basis into the twistor basis.
The change of basis  can be made using the spinor representations of $\mathfrak{so}(2,d)$.
We shall first find the spinor representations of Lorentz subalgebra $\mathfrak{so}(1,d-1)$
and extend them to $\mathfrak{so}(2,d)$.

Let us begin with the $d=3$ case, $\mathfrak{so}(1,2)\subset \mathfrak{gl}(2,\mathbb R)$, 
where the gamma matrices are indeed all real and given by
\be
	\overset{\sst (1,2)}{\g}_\m=\t_\mu\,,
\ee
with
\be
    \t_0=i\,\s_2=
     \begin{pmatrix}
    0 & 1 \\
    -1 & 0
    \end{pmatrix},
    \qquad 
    \t_1=\s_1=
     \begin{pmatrix}
    0 & 1 \\
    1 & 0
    \end{pmatrix},
    \qquad
    \t_2=\s_3=
     \begin{pmatrix}
    1 & 0 \\
    0 & -1
    \end{pmatrix}.
    \label{eq:tau matrices}
\ee 
The $\t_\mu$ matrices satisfy the (anti-)commutation relations,
\be
	[\t_\mu,\t_\nu]=2\,\epsilon_{\m\nu\rho}{}\,\t^\rho\,,
	\qquad
	\{\t_\mu,\t_\nu\}=2\,\eta_{\m\n}\,,
\ee
with the conventions,
\be
	\epsilon_{012}=1\,,\qquad  \eta_{\m\n}={\rm diag}(-1,1,1)\,.
\ee
This choice of basis can be extended to the Clifford algebra $\mathfrak{Cl}(2,3)\supset \mathfrak{b}(\O_{\sst (4)},\mathbb R)$  as
\be
	\overset{\sst (2,3)}{\g}_{0'}=\hat\t_0\otimes 1_{\sst(2)}\,,\qquad
	\overset{\sst (2,3)}{\g}_\m=\hat\t_2\otimes \t_\mu\,,\qquad 
	\overset{\sst (2,3)}{\g}_{3}=\hat\t_1\otimes 1_{\sst(2)}\,,
\ee
while keeping reality of the gamma matrices,
or to $\mathfrak{Cl}(1,5)\supset \mathfrak{sl}(2,\mathbb H)$ with the gamma matrices,
\be
	\overset{\sst (1,5)}{\g}_\m= \t_\mu\otimes \check\t_1\,,\qquad 
	\overset{\sst (1,5)}{\g}_{\o}=\mathsf{q}_{\o-3}\,1_{\sst(2)}\otimes\check \t_0 \quad [\o=4,5,6]\,,
\ee
where $(\mathsf{q}_1,\mathsf{q}_2,\mathsf{q}_3)=(\mathsf i,\mathsf j,\mathsf k)$ are the quaternion basis.
The first element $\mathsf i$ can serve as $\sqrt{-1}$ for the subalgebra $\mathfrak{Cl}(1,3)\supset \mathfrak{sl}(2,\mathbb C)$.
We introduced different notations $\hat\t_\mu=\check\t_\mu=\t_\mu$ to make 
the inclusion structure more clear.
Finally the largest Clifford algebra $\mathfrak{Cl}(2,6)\supset \mathfrak{b}(\O_{\sst (4)},\mathbb H)$ 
is realized with the gamma matrices,
\ba
	&\overset{\sst (2,6)}{\g}_{0'}=\hat\t_0\otimes 1_{\sst(2)} \otimes\check\t_1\,,\qquad
	&\overset{\sst (2,6)}{\g}_\m=\hat\t_2\otimes \t_\mu\otimes\check \t_1\,,\nn
	&\overset{\sst (2,6)}{\g}_{3}=\hat\t_1\otimes 1_{\sst(2)} \otimes \check \t_1\,,\qquad 
	&\overset{\sst (2,6)}{\g}_{\o}=\mathsf q_{\o-3}\,\hat 1_{\sst(2)} \otimes 1_{\sst(2)} \otimes \check \t_0\,.
\ea
These gamma matrices are $8\times 8$ with $\mathbb H$ entries, but 
the associated Lie algebra generators are all proportional to $\check 1_{\sst(2)}$ or $\check\t_2$ 
(no dependence on $\check\t_0$ and $\check\t_1$), 
and hence the last tensor product entry with $\check{\cdot}$ can be dropped.
In the end, we identified the 
spinor representations of all $\mathfrak{so}(2,6)\cong \mathfrak{b}(\O_{\sst (4)},\mathbb H)$ generators.
For the realization of $\mathfrak{b}(\O_{\sst (4)},\mathbb H)$ with
the standard form of $\O_{\sst (4)}$, 
\be
	\O_{\sst (4)}=\hat \t_0\otimes 1_{\sst(2)} =
	\begin{pmatrix}
    0 & 1_{\sst (2)} \\
    -1_{\sst (2)} & 0
    \end{pmatrix},
\ee 
we have to intertwine the representations as
\be
	J_{AB}=Q^{-1} \,\frac{[\g_A,\g_B]}4\, Q\qquad [A,B=0',\mu,\o]\,,
\ee 
where the intertwiner $Q$ is given by
\be
	Q=\begin{pmatrix} 1_{\sst(2)} & 0 \\ 0 & \t_0 \end{pmatrix}\,.
\ee
See Appendix \ref{sec:ads dic} for
the explicit component expressions
of $J_{AB}$.
The dual basis $\cJ^{AB}$ of $\mathfrak{b}(\O_{\sst (4)},\mathbb F)^*$ can be also
represented by $4\times 4$ matrices as
\be
	\cJ^{AB}=J^{BA}\,,
\ee
and satisfy
\be \label{eq:pairing condition}
	\la \cJ^{AB},J_{CD}\ra
	=\tr(\cJ^{AB}\,J_{CD})=\delta^{[A}_{[C}\,\delta^{B]}_{D]}\,.
\ee
In the following, 
we find simple 
matrix expressions
of various coadjoint vectors $\phi$,
by making use of the matrix representation
of $\cJ^{AB}$ introduced above.
We then diagonalize the matrices $\phi$ to identify the dual groups
and the corresponding dual coadjoint vectors $\tilde\phi$. 
The matrix representation of  each $\tilde\phi$ 
determines the worldline action
of the corresponding twistor particle:
it is enough to put the data of $\tilde\phi$ into 
the general expression \eqref{AdS twistor action}
of the twistor action.

Before enlisting our results, let us remark that 
we assume none of the parameters
of $\phi$ or $\tilde\phi$ vanish.
To address such limiting case, we perform 
separate analysis. 
Moreover, 
we limit ourselves to the coadjoint orbits with at most two independent labels, for simplicity.
This will exclude for instance a large class of mixed symmetry spin particles.

\subsection{Particles with time-like momenta}

We begin with the particle species whose
representative momenta are time-like.

\begin{itemize}

\item
Let us consider first
the case
of massive scalar particles.
The corresponding coadjoint orbit is given by the representative vector,
\be
    \f= m\,\cJ^{0'0}\,,
\ee
and it can be written in the $4\times4$ matrix form as
\be
    \f\,\O_{\sst(4)}^{\ -1}=\frac m2
    \begin{pmatrix}
        1_{\sst(2)} & 0 \\
        0 & 1_{\sst(2)}
    \end{pmatrix}.
\ee
From the above, we identify that the dual group $\tilde G$ is $B(\h_{\sst (4)},\mathbb F)$,
and the dual coadjoint orbit is given by the coadjoint element,
\be
    \h_{\sst(4)}\,\tilde\f=
    \frac m2\,\begin{pmatrix}
        0 & 1_{\sst(2)} \\
        -1_{\sst(2)} & 0
    \end{pmatrix},
    \label{scalar dual coadj}
\ee
which immediately specify the 
twistor particle action through \eqref{AdS twistor action}.
We remind the reader that
the constraints analysis of particle action can be universally replaced by
the analysis of the dual stabiliser. 
For that, it is useful 
to express the matrix form 
\eqref{scalar dual coadj} of the coadjoint element 
in terms of the dual basis $\cR^{IJ}$ as
\be
	\tilde\phi  = \frac{|m|}2\,(\cR^{[13]}+\cR^{[24]})\,.
\ee 
Then, it becomes fairly straightforward to
analyze the stabiliser $\tilde{\mathfrak{g}}_{\tilde\f}$.
We find that $\tilde{\mathfrak{g}}_{\tilde\f}$ is
generated by
\be
    R_{[13]}+R_{[24]}\,,\quad 
     R_{[13]}-R_{[24]}\,,\quad R_{[12]}+R_{[34]}\,,\quad R_{[14]}+R_{[23]}\,,
\ee
\be
     \mathsf{q}_\a\,(R_{11}+R_{33})\,,\quad
     \mathsf{q}_\a\,(R_{22}+R_{44})\,,\quad
     \mathsf{q}_\a\,(R_{(12)}+R_{(34)})\,,\quad 
     \mathsf{q}_\a\,(R_{(14)}-R_{(23)})\,,
\ee
where $(\mathsf q_1, \mathsf q_2, \mathsf q_3)=
(\mathsf i, \mathsf j, \mathsf k)$ 
are the quaternion basis, 
and hence the generators containing $\mathsf q_\a$ are all absent for $\mathbb F=\mathbb R$, 
and only those with $\mathsf q_1=\mathsf i$ are present for $\mathbb F=\mathbb C$.
Note that the stabiliser algebra $\tilde{\mathfrak{g}}_{\tilde\f}$ is isomorphic
to $\mathfrak{u}(2)$, $\mathfrak{u}(2)\oplus
\mathfrak{u}(2)$ and $\mathfrak{u}(4)$
for $\mathbb F=\mathbb R$, $\mathbb C$ and $\mathbb H$, respectively.
Since ${\rm dim}(\tilde{\mathfrak{g}}_{\tilde\f})=
    4\,|\mathbb F|$,
we find the dimension of the coadjoint orbit, according to \eqref{dim orbit}, as
\be
	{\rm dim}(\cO_\phi^G) = 2\,|\mathbb F|+4\,.
 \label{dim scalar}
\ee
The phase space dimension of  a scalar particle in AdS$_{d}$ is $2(d-1)$,
that is $6, 8$ and $12$ for $d=4,5$ and $7$, respectively,
and hence matches the dimensions \eqref{dim scalar} 
for $\mathbb F=\mathbb R, \mathbb C$ and $\mathbb H$.

\item 

Let us move to the case of spinning particles.
The coadjoint orbit with
    time-like momentum
    and space-like spin
    is given by
\be
	\phi =  m\,\cJ^{0'0}+s\,\cJ^{12} \,,
\ee
or, in $4\times 4$ matrix form, by
\be
    	\phi\,\O_{\sst (4)}^{\ -1}=\frac12 \begin{pmatrix}
	m\,1_{\sst(2)} & s\, \t_0 \\
	-s\, \t_0 & m\,1_{\sst(2)}
	\end{pmatrix}.
\ee    
We can diagonalize this matrix to identify the dual group $\tilde G$ 
and the corresponding dual orbit.
At this step, we find three sub-cases
to which we accord
different particle interpretations.

\begin{itemize}

\item
The case with $|m| > |s|$
is interpreted as a massive spinning particle.
The dual group $\tilde G$ is 
$B(\h_{\sst (4)},\mathbb F)$, the same as the scalar case,
while the dual coadjoint orbit is given by 
the coadjoint element,
\be
    \tilde\f=(\eta_{\sst (4)})^{-1} 
    \frac12\, \begin{pmatrix}
    |m+s|\,\t_0 & 0 \\
    0 & |m-s|\,\t_0
    \end{pmatrix}
    = \frac{|m+s|}2\,\cR^{[12]}+\frac{|m-s|}2\,\cR^{[34]}\,.
    \label{MS m}
\ee
The dual stabiliser is 
\be 
  \tilde{\mathfrak{g}}_{\tilde \f} \cong \mathfrak{u}(1)_{R_{[12]}} \oplus 
    \hat{\mathbb F}_{R_{11}+R_{22}} \oplus
    \mathfrak{u}(1)_{R_{[34]}} \oplus
 \hat{\mathbb F}_{R_{33}+R_{44}} \,,
 \label{massive twist dual stal}
\ee
where $\hat{\mathbb F}_{X}$ is the Lie algebra 
$\hat \R=\mathfrak{o}(1)=0$,
$\hat \C=\mathfrak{u}(1)$
and $\hat{\mathbb H}
=\mathfrak{sp}(1)$ generated by $\mathsf{q}_\a\,X$\,.
In this spinning case,
the dimension of dual stabiliser is reduced to
${\rm dim}( \tilde{\mathfrak{g}}_{\tilde \f} )=2\,|\mathbb F|$\,.
Consequently, the dimension of the coadjoint orbit is increased to 
\be
	{\rm dim}(\cO_\phi^G) = 4\,|\mathbb F|+4\,.
	\label{massive twist orbit dim}
\ee
Since
the phase space dimension of a massive spinning particle in AdS$_{d}$ is $2(2d-4)$,
it matches 
the dimension \eqref{massive twist orbit dim}
in 4, 5 and 7 dimensions.

\item

The case with $|m|=|s|$
is interpreted as a massless spinning particle. 
In this case, the dual group $\tilde G$ reduces to 
 $B(\h_{\sst (2)},\mathbb F)$
with
\be 
   \tilde\f=
    (\eta_{\sst (2)})^{-1}\,|s|\,\t_0= |s|\,\cR^{[12]}\,.
\ee
The dual stabiliser is also the half of the massive case,
\be
  \tilde{\mathfrak{g}}_{\tilde \f} \cong \mathfrak{u}(1)_{R_{[12]}}
  \oplus \hat{\mathbb F}_{R_{11}+R_{22}} \,,
\ee
with ${\rm dim}( \tilde{\mathfrak{g}}_{\tilde \f} )=|\mathbb F|$\,.
The dimension of the coadjoint orbit is 
\be
	{\rm dim}(\cO_\phi^G) = 4\,|\mathbb F|+2\,,
	\label{massless twist orbit dim}
\ee
and matches the phase space dimension of a massless spinning particle in AdS$_{d}$,
that is $2(2d-5)$, in 4, 5 and 7 dimensions.

\item
The case with $|m|<|s|$ 
can be better interpreted as a particle of mass $s$ and time-like spin $m$
living in BdS.
The dual group is $B(\h_{\sst (2,2)},\mathbb F)$,
while the dual orbit is given by the same coadjoint element \eqref{MS m}
with the same dual stabiliser \eqref{massive twist dual stal}.

\end{itemize}
\end{itemize}

\subsection{Particles with light-like momenta}

Let us move to the case where
the momentum of particle is null, namely light-like.
As we will show shortly, this case includes
the massless scalar and continuous spin particles,
together with the massless limit of massive tachyons.

\begin{itemize}

\item 
The coadjoint orbit of the massless scalar particle
is given by
\be
    \f\,\O_{\sst(4)}^{\ -1}=
    E\, \cJ^{0'+}\, \O^{\ -1}_{\sst(4)}
    =\frac E4 \begin{pmatrix}
    1_{\sst(2)} & 1_{\sst(2)} \\
    1_{\sst(2)} & 1_{\sst(2)}
    \end{pmatrix},
\ee
with $\cJ^{0'+}=\frac12(\cJ^{0'0}+\cJ^{0'3})$.
The dual group is $\tilde G=B(\h_{\sst (2)},\mathbb{F})$\,,
and the dual coadjoint orbit is 
the trivial orbit,
\be
    \tilde \f=0\,.
\ee
In this case, the dual stabiliser coincides with the dual group itself, and 
hence ${\rm dim}(\tilde g_{\tilde\phi})=3|\mathbb F|-2$.
Consequently, the dimension of the 
coadjoint orbit,
\be
	{\rm dim}(\cO_\phi^G) = 2\,|\mathbb F|+4\,,
\ee
coincides with the phase space dimension of a scalar particle in AdS$_{|\mathbb F|+3}$.

\item 
The coadjoint orbit with light-like momentum
 and space-like spin,
 corresponding to the endpoint of the massive tachyonic particle, is given by
 \be
	   \f\,\O_{\sst (4)}^{\ -1}  =
	   (E\, \cJ^{0'+}+s\,\cJ^{12})\,\O_{\sst (4)}^{\ -1}  
	   =\frac12 \begin{pmatrix}
    \frac E2\,1_{\sst(2)} & \frac E2\,1_{\sst(2)}+s\,\t_0 \\
    \frac E2\,1_{\sst(2)}-s\,\t_0 & \frac E2\,1_{\sst(2)}
  \end{pmatrix}.
 \ee 
The dual group is $\tilde G= B(\h_{\sst (2,2)},\mathbb{F})$,
and the dual coadjoint orbit is given by
\ba
	\tilde\f \eq     (\eta_{\sst (2,2)})^{\!-1}\,\frac14
   \begin{pmatrix}
   -(2\,|s|-\e)\,\t_0 & -\e\,1_{\sst(2)} \\
   \e\,1_{\sst(2)} & (2\,|s|+\e)\,\t_0
   \end{pmatrix}\nn
   \eq 
    -|s|\left(\cR^{[+'+]}+\cR^{[-'-]}\right)
    +\e\,\cR^{[+'-]}\,.
\ea
The dual stabiliser is
 \ba 
    \tilde{\mathfrak g}_{\tilde\f}&\cong&
    \R_{R_{[+'+]}+R_{[-'-]}-2R_{[+'-]}}
    \oplus \mathfrak{u}(1)_{R_{[+'+]}+R_{[-'-]}} \nn
    \qquad &&
    \oplus\, \hat{\mathbb F}_{R_{(+'-')}+R_{(+-)}}
    \oplus \hat{\mathbb F}_{R_{+'+'}-R_{-'-'}-R_{++}+R_{--}}\,,
\ea 
where the second line is trivial for $\mathbb F=\mathbb R$,
and isomorphic to $\mathfrak{u}(1)\oplus\mathfrak{u}(1)$
and $\mathfrak{sl}(2,\C)$
for $\mathbb{F}=\C$ and $\mathbb H$, respectively. 
Here,  the lightcone indices, $\cV^{\pm'}=\frac12(\cV^{0'}\pm \cV^2)$ and 
$\cV^{\pm}=\frac12(\cV^{0}\pm\cV^1)$, are used also for the dual group indices.
Note that $\e\cong \l\,\e$ for any $\l \in \R_{\neq 0}$: it can be rescaled using the mutual  boost
in the $+/-$ and $+'/-'$ coordinates.

The presence of $\e$ parameters shows that the corresponding orbit is 
the special semi-simple orbits discussed in Section 7.2 of \cite{Basile:2023vyg}:
when the dual group is regarded as the spacetime symmetry,
the dual orbit corresponds to the particle with entangled mass and spin, 
and the parameter $\e$ is associated with a nilpotent coadjoint element.

\item 
The coadjoint orbit with light-like momentum
 and light-like spin,
 corresponding 
 to the continuous spin particle, is given by
\be
	  \phi\,\O_{\sst (4)}^{\ -1}=
	  \left(E\,\cJ^{0'+}+\e \,\cJ^{-2}\right) \O_{\sst (4)}^{\ -1}=
	  \frac14 
    \begin{pmatrix}
    E\,1_{\sst(2)} + \e\,\t_1 & E\,1_{\sst(2)} - \e\,\t_1 \\
    E\,1_{\sst(2)}-\e\,\t_1 & E\,1_{\sst(2)} + \e\,\t_1
    \end{pmatrix}.
\ee
In this case, the dual group $\tilde G$ is $B(\h_{\sst (1,3)},\mathbb F)$, and the dual orbit is given by
\be 
 	\tilde\f=(  \eta_{\sst (1,3)})^{-1}\,
    \frac{\r}{2}\,
    \begin{pmatrix}
    \t_0 & 0  \\
     0 & -\t_0
    \end{pmatrix}
    =\frac{\r}2\,(\cR^{[01]}-\cR^{[23]})\,,
    \label{contispin}
\ee 
with the stabiliser,
\be
   \tilde{\mathfrak{g}}_{\tilde \f} \cong 
   \R_{R_{[01]}}\oplus  
     \hat{\mathbb F}_{R_{00}-R_{11}}\oplus
   \mathfrak{u}(1)_{R_{[23]}} \oplus
    \hat{\mathbb F}_{R_{22}+R_{33}}\,.
\ee
Here, $\rho=\sqrt{|E\,\e|}$ is the genuine label of the dual coadjoint orbit,
which cannot be freely rescaled.

\end{itemize}

\subsection{Particles with space-like momenta}
The particles with space-like momenta, namely tachyons, have infinitely many degrees of freedom\footnote{Quantizations of these particles will lead to irreps
which are realized by infinite component fields, like in the continuous spin case.}
except for the scalar case. 
For the spinning cases, there are several different kinds.

\begin{itemize}

\item 
The coadjoint orbit of the tachyonic scalar particle
is given by
\be
    \f\,\O_{\sst(4)}^{\ -1}=
    \m\, \cJ^{0'1}\,\O^{\ -1}_{\sst(4)}
    =\frac \m2 \begin{pmatrix}
    -\t_2 & 0 \\
    0 & \t_2
    \end{pmatrix}.
\ee
The dual group is $\tilde G=B(\h_{\sst (2,2)},\mathbb{F})$
and the dual orbit is given by
\be
    \tilde \f=
    (\h_{\sst (2,2)})^{-1}\,
    \frac{|\m|}2\,
    \begin{pmatrix}
        0 & \t_2 \\
        -\t_2 & 0
    \end{pmatrix}
    =\frac{|\m|}2\,
    (\cR^{[0'1]}-\cR^{[02]})\,,
\ee
with the stabiliser $\tilde{\mathfrak{g}}_{\tilde\f}
\cong \mathfrak{gl}(2,\mathbb F)$
generated by 
\be
    R_{[0'1]}-R_{[02]}\,, \quad 
    R_{[0'0]}-R_{[12]}\,,\quad 
    R_{[0'1]}+R_{[02]}\,,\quad
    R_{[0'2]}-R_{[01]}\,,
\ee
\be
    \mathsf{q}_\a\,(R_{0'0'}-R_{11})\,,\quad
    \mathsf{q}_\a\,(R_{00}-R_{22})\,,\quad
    \mathsf{q}_\a \,(R_{(0'0)}-R_{(12)})\,,\quad
    \mathsf{q}_\a \,(R_{(0'2)}+R_{(01)})\,.
\ee

\item The coadjoint orbit with space-like momentum
 and time-like spin is given by
\be
     \phi\,\O_{\sst (4)}^{\ -1}
	=\left(\mu\,\cJ^{0'1}+\nu\,\cJ^{02}\right)\O_{\sst (4)}^{\ -1}
	 = \frac12 \begin{pmatrix}
   -\m\,\t_2 & -\n\,\t_1 \\
    -\n\,\t_1 & \m\,\t_2
    \end{pmatrix}.
    \label{mu nu tachyon}
\ee
Up to conjugation, we can take $\mu\ge \nu>0$.
\begin{itemize}
\item
For $\mu> \nu$, the dual group $\tilde G$ is $B(\h_{\sst(2,2)},\mathbb F)$,
and the dual coadjoint orbit is given by
\ba 
    \tilde\f\eq (\eta_{\sst (2,2)})^{-1}\,\frac12
    \begin{pmatrix}
   0  &  \m\,\t_0+\n\,\t_1 \\
   \mu\,\t_0-\n\,\t_1 & 0 
    \end{pmatrix} \nn
\eq \frac{\m+\n}2\,\cR^{[0'2]}-\frac{\m-\n}2\,\cR^{[01]}\,,
    \label{tachyonT}
\ea
with the stabiliser,
\be
    \tilde{\mathfrak{g}}_{\tilde\f}\cong
    \mathbb{R}_{R_{[01]}}
     \oplus \hat{\mathbb{F}}_{R_{00}-R_{11}}
     \oplus
    \mathbb{R}_{R_{[0'2]}}
    \oplus \hat{\mathbb{F}}_{R_{0'0'}-R_{22}}
   \,.
\ee
\item
For short tachyon with $\mu = \nu$,
the dual group $\tilde G$ is $B(\h_{\sst (1,1)},\mathbb F)$,
and the dual coadjoint orbit is given by
\be 
   \tilde\f=( \eta_{\sst (1,1)})^{-1}\,
    \n\,\t_0=\n\,\cR^{[01]}\,,
    \label{short tachyon}
\ee
with the stabiliser,
\be
    \tilde{\mathfrak{g}}_{\tilde\f}\cong 
    \mathbb{R}_{R_{[01]}}
    \oplus \hat{\mathbb{F}}_{R_{00}-R_{11}}\,.
\ee
\end{itemize}

\item The coadjoint orbit with space-like momentum
 and space-like spin is given by
\be
    \phi\,\O_{\sst (4)}^{\ -1}=
(\mu\,\cJ^{0'1}+s\,\cJ^{23})\,\O_{\sst (4)}^{\ -1}
    = \frac12 \begin{pmatrix}
   -\m\,\t_2+s\,\t_1 & 0\\
   0 & \m\,\t_2+s\,\t_1
    \end{pmatrix}.
\ee
The dual group $\tilde G$ is
$ B(\h_{\sst (2,2)},\mathbb F)$,
and the dual coadjoint orbit is given by
\be
	\tilde\f = 
	(\h_{\sst (2,2)})^{-1}\,\frac12
	\begin{pmatrix}
	|s|\,\t_0 & |\m|\,\t_1 \\
	-|\m|\,\t_1 & |s|\,\t_0
	\end{pmatrix}
	=
            \frac {|s|}2\left(\cR^{[0'0]}+\,\cR^{[12]}\right)
            +\frac{|\m|}2\left(\cR^{[0'2]}+\cR^{[01]}\right)\,,
            \label{tachyonS}
\ee
with the stabiliser,
\ba
       \tilde{\mathfrak g}_{\tilde\f}& \cong &
       \mathfrak{u}(1)_{R_{[0'0]}+R_{[12]}}\oplus \R_{R_{[0'2]}+R_{[01]}} \nn
       &&
       \oplus\, \hat{\mathbb{F}}_{R_{0'0'}+R_{00}-R_{11}-R_{22}}
       \oplus \hat{\mathbb{F}}_{R_{(0'1)}-R_{(02)}}\,.
\ea 
Note that the second line
is trivial for $\mathbb F=\mathbb R$
and isomorphic to 
$\mathfrak{u}(1)\oplus \mathbb R$ and $\mathfrak{sl}(2,\C)$ for
$\mathbb F=\C$ and $\mathbb H$, respectively. 
The dual coadjoint vector $\tilde\f$ can be interpreted as
the coadjoint vector of the one-parameter family of continuous spinning particle
when the dual group $\tilde G$ is regarded as the spacetime symmetry.

\item 
The coadjoint orbit with space-like momentum
 and light-like spin is given by
 \be
   \f\,\O_{\sst (4)}^{\ -1}=
(\m\, \cJ^{0'1}+\e\,\cJ^{2+})\,\O_{\sst (4)}^{\ -1}
   = \frac{1}{2}
   \begin{pmatrix}
   0 & 0 & \m & -\e \\
   0 & 0 & 0 & \m \\
   \m & 0 & 0 & 0 \\
   -\e & \m & 0 & 0
   \end{pmatrix},
 \ee
with $\cJ^{2\pm}=\frac12(\cJ^{20}\pm\cJ^{23})$.
The dual group $\tilde G$ is $B(\h_{\sst (2,2)},\mathbb F)$,
and the dual orbit is given by
\ba
        \tilde\phi
        \eq (\h_{\sst (2,2)})^{-1}\,\frac{1}{4}
        \begin{pmatrix}
        \ve\,\t_0 & 2|\m|\,\t_1+\ve\,\t_0 \\
        -2|\m|\,\t_1+\ve\,\t_0 & \ve\,\t_0
        \end{pmatrix} \nn
       \eq
	  |\m|\left(\cR^{[-'+]}-\cR^{[+'-]} \right)
  	+\ve\,\cR^{[+'+]}\,,
   \label{tachyonN}
\ea
with the stabiliser,
\ba 
    \tilde{\mathfrak{g}}_{\tilde\f} &\cong &
    \R_{R_{[+'+]}} \oplus
    \R_{R_{[+'-]}-R_{[-'+]}} \nn 
    &&\oplus\,
    \hat{\mathbb{F}}_{R_{+'+'}-R_{++}} \oplus
    \hat{\mathbb{F}}_{R_{(+'-')}+R_{(+-)}}\,.
\ea 
The second line is trivial for $\mathbb{F}=\R$ and
isomorphic to $\mathbb R\oplus \mathbb R$ and 
$\mathfrak{iso}(3)$ 
for $\mathbb F=\C$ and $\mathbb H$, respectively.
Here, $\cV^\pm=\frac12(\cV^0\pm \cV^2)$
and $\cV^{\pm'}=\frac12(\cV^{0'}\pm \cV^1)$,
and $\ve\cong \l\,\ve$ for any $\l\in\mathbb R_{\neq 0}$.

The dual coadjoint vector $\tilde\f$ can be understood as 
the coadjoint vector of the entangled short tachyon
if the dual group is regarded as the spacetime symmetry.

\end{itemize}

\subsection{Particles
with entangled mass and spin}

Next, we consider the cases that
we referred to as \emph{entangled mass and spin}, in the sense
that the representative coadjoint vector is given by the sum of two coadjoint elements in a subtle relation.
For detailed discussion, we refer to our previous work \cite{Basile:2023vyg}.

\begin{itemize}

\item The coadjoint orbit of one parameter family of continuous spin particle  
is given by
\ba
	\phi\,\O_{\sst (4)}^{\ -1}\eq 
    \left(s\,(\cJ^{0'0}+\cJ^{12})+\n\,(\cJ^{0'1}-\cJ^{02})\right)\O_{\sst (4)}^{\ -1} \nn
    \eq
        \frac12
    \begin{pmatrix}
      s\,1_{\sst(2)}-\n\,\t_2 & s\,\t_0+\n\,\t_1 \\
      -s\,\t_0+ \n\,\t_1 & s\,1_{\sst(2)}+\n\,\t_2
    \end{pmatrix}.
\ea
The dual group $\tilde G$ is $B(\h_{\sst (1,3)},\mathbb{F})$,
and the dual coadjoint orbit is given by
\be
  \tilde\f = (\h_{\sst(1,3)})^{-1}
  \begin{pmatrix}
   -|\n|\,\t_0 & 0 \\
   0 & |s|\,\t_0
  \end{pmatrix}= |\n|\,\cR^{[01]}-|s|\,\cR^{[23]}\,,
  \label{conti}
\ee
with the stabiliser,
\be
        \tilde{\mathfrak{g}}_{\tilde\f}\cong
        \R_{R_{[01]}}\oplus
        \hat{\mathbb{F}}_{R_{00}-R_{11}}\oplus
        \mathfrak{u}(1)_{R_{[23]}}\oplus
        \hat{\mathbb{F}}_{R_{22}-R_{33}}\,.
\ee
 Depending on the value of $\nu$,
the cases
with $\nu>s$, $\nu<s$ and $\nu=s$
may be interpreted 
as massive, tachyonic and massless
 particles of continuous spin,
 and the last case coincides
 with the usual continuous spin particle treated in \eqref{contispin}.
 
\item The coadjoint orbit of entangled massless particle 
is given by
\ba
	\phi\,\O_{\sst (4)}^{\ -1}\eq 
    \left(s\,(\cJ^{0'0}+\cJ^{12})+\e\,(\cJ^{0'0}+\cJ^{0'1}-\cJ^{12}-\cJ^{02})\right)\O_{\sst (4)}^{\ -1} \nn
    \eq \begin{pmatrix}
      \frac s2 & 0 & 0 & \frac s2 \\
      0 & \frac s2+\e & -\frac s2+\e & 0 \\
      0 & -\frac s2+\e & \frac s2+\e & 0 \\
      \frac s2 & 0 & 0 & \frac s2
    \end{pmatrix}.
    \label{entangled}
\ea
The dual group $\tilde G$ is $B(\h_{\sst (3)},\mathbb{F})$
for $s\,\e>0$ 
and $B(\h_{\sst (1,2)},\mathbb{F})$
for $s\,\e<0$.
In both cases, the dual coadjoint orbit is given by
\be
  \tilde \f =(\h_{\sst (p,q)})^{-1}\,
  \begin{pmatrix}
    0 & |s| & 0 \\
    -|s| & 0 & 0 \\
    0 & 0 & 0
  \end{pmatrix}= -|s|\,\cR^{[12]}\,,
  \label{entangleSN}
\ee
with the stabiliser,
\be
       \tilde{\mathfrak{g}}_{\tilde\f}\cong
       \mathfrak{u}(1)_{R_{[12]}}\oplus
       \hat{\mathbb{F}}_{R_{11}+R_{22}}\oplus
       \hat{\mathbb{F}}_{R_{\bullet\bullet}}\,.
\ee
Here, the index $\bullet$ indicates the spatial direction for $s\,\e>0$ case,
and the temporal direction for $s\,\e<0$

These cases correspond to the large remnant orbits of massive spinning orbits in the massless limit
approaching from $m>s$ and $m<s$, respectively.

\item The coadjoint orbit of entangled short tachyon  
is given by
\ba
	\phi\,\O_{\sst (4)}^{\ -1} \eq 
   \left(\n\,(\cJ^{0'2}+\cJ^{01})+\e\,(\cJ^{0'0}+\cJ^{0'2}-\cJ^{21}-\cJ^{01}) \right) \O_{\sst (4)}^{\ -1} \nn 
    \eq \frac12
    \begin{pmatrix}
    \e\,1_{\sst(2)}+(\e+\n)\,\t_1 & \e\,\t_0-(\e-\n)\,\t_2 \\
    -\e\,\t_0-(\e-\n)\,\t_2 & \e\,1_{\sst(2)}-(\e+\n)\,\t_1
    \end{pmatrix}.
\ea
The dual group $\tilde G$ is $B(\h_{\sst (1,2)},\mathbb{F})$.
The dual coadjoint orbit is given by
\be
	\tilde \f=(\h_{\sst (p,q)})^{-1}\,
	\begin{pmatrix}
	 0 & |\n| & 0 \\
	  -|\n| & 0 & 0 \\
	  0 & 0 & 0
	\end{pmatrix}
	= -|\n|\,\cR^{[01]}\,,
    \label{entangleTN}
\ee
with the dual stabiliser,
\be
       \tilde{\mathfrak{g}}_{\tilde\f}\cong
       \R_{R_{[01]}}\oplus
       \hat{\mathbb{F}}_{R_{00}-R_{11}}\oplus
       \hat{\mathbb{F}}_{R_{22}}\,.
\ee
This case corresponds to the large remnant orbit of a tachyonic spinning orbit in the massless limit.

\end{itemize}

\subsection{Particles in BdS spacetime}

Let us move to consider
the particles in bitemporal AdS (BdS) spacetime defined as the hypersurface $\eta_{AB}\,X^A\,X^B=+L^2$ with
the metric $\eta_{AB}={\rm diag}(--+\cdots+)$.
BdS was identified originally
in the vectorial description 
where the hypersurface equation arise
as a constraint.
The BdS particles have many sub cases
including the massive spinning particle
with $|m|<|s|$ treated before.

\begin{itemize}
    \item The coadjoint orbit 
    of the BdS ``massive'' scalar 
    is given by
\be
	\phi\,\O_{\sst (4)}^{\ -1}=
    m\,\cJ^{12}\,\O_{\sst (4)}^{\ -1} = 
    \frac m2\begin{pmatrix}
    0 & \t_0 \\
    -\t_0 & 0
    \end{pmatrix}.
\ee
The dual group is
$\tilde G=B(\h_{\sst (2,2)},\mathbb F)$,
and the dual coadjoint orbit is given by
\be
    \tilde \f=(\h_{\sst (2,2)})^{-1}\,
    \frac{|m|}2\,
    \begin{pmatrix}
        \t_0 & 0 \\
        0 & \t_0
    \end{pmatrix}
    =\frac{|m|}2(\cR^{[0'0]}+\cR^{[12]})\,.
\ee
The dual stabiliser $\tilde{\mathfrak g}_{\tilde\phi}$ generated by 
\be
    R_{[0'0]}+R_{[12]}\,,\quad 
    R_{[0'0]}-R_{[12]}\,,\quad
    R_{[0'1]}-R_{[02]}\,,\quad
    R_{[0'2]}+R_{[01]}\,,
\ee
\be
    \mathsf{q}\,(R_{0'0'}+R_{00})\,,\quad
    \mathsf{q}_\a\,(R_{11}+R_{22})\,,\quad
    \mathsf{q}_\a\,(R_{(0'1)}-R_{(02)})\,,\quad
    \mathsf{q}_\a\,(R_{(0'2)}+R_{(01)})\,,
\ee
is isomorphic to $\mathfrak{u}(1,1)$,
$\mathfrak{u}(1,1)\oplus\mathfrak{u}(1,1)$
and $\mathfrak{u}(2,2)$ 
for $\mathbb F=\R, \C$ and $\mathbb{H}$.

    \item The coadjoint orbit 
    of the BdS ``massless'' scalar 
    is given by
\be
	\phi\,\O_{\sst (4)}^{\ -1}=
    E\,\cJ^{1+}\,\O_{\sst (4)}^{\ -1} = 
    \frac E2\begin{pmatrix}
    0 & \t_0-\t_2 \\
    -\t_0+\t_2 & 0
    \end{pmatrix},
\ee
where $\cJ^{1+}=\frac12(\cJ^{10}+\cJ^{13})$.
The dual group is
$\tilde G=B(\h_{\sst (1,1)},\mathbb F)$
and its dual coadjoint orbit is the trivial orbit,
\be
    \tilde\f=0\,.
\ee
\end{itemize}

\medskip

The BdS spinning particles exist only in AdS$_{d>4}$,
and hence we restrict $\mathbb{F}$ to $\C$ and $\mathbb{H}$.

\begin{itemize}

\item The coadjoint orbit of BdS spinning particle
with space-like momentum and space-like spin
is given by
\be
	\phi\,\O_{\sst (4)}^{\ -1}=
    (m\,\cJ^{12}+s\,\cJ^{34})\,\O_{\sst (4)}^{\ -1} = 
    \frac12\begin{pmatrix}
    \mathsf{i}\,s\,\t_0 & m\,\t_0 \\
    -m\,\t_0 & \mathsf{i}\,s\,\t_0
    \end{pmatrix}.
\ee
\begin{itemize}

\item In the ``massive'' case with $m\neq s$, the dual group 
is $B(\h_{\sst (2,2)},\mathbb{F})$,
and the dual coadjoint orbit is given by
\ba
  \tilde\f \eq
  (\h_{\sst (2,2)})^{-1}\,\frac{\mathsf{i}}{2}\,
  {\rm diag }\,(|m+s|\,,-|m-s|\,,-|m-s|\,,|m+s|) \nn
  \eq
      \frac{\mathsf{i}}{2}\,
      \left(|m+s|(\cR^{0'0'}+\cR^{22})
      -|m-s|\,(\cR^{00}+\cR^{11}) \right).
      \label{BdS}
\ea
For $\mathbb{F}=\C$, the stabiliser  is
\be
     \tilde{\mathfrak{g}}_{\tilde\f}\cong
     \mathfrak{u}(1)_{\mathsf{i}\,R_{0'0'}}\oplus
     \mathfrak{u}(1)_{\mathsf{i}\,R_{00}}\oplus
     \mathfrak{u}(1)_{\mathsf{i}\,R_{11}}\oplus
     \mathfrak{u}(1)_{\mathsf{i}\,R_{22}}\,,
\ee
while for $\mathbb{F}=\mathbb{H}$, the stabiliser  is
\be
     \tilde{\mathfrak{g}}_{\tilde\f}\cong
     \mathfrak{u}(1,1)_{0'2}\oplus
      \mathfrak{u}(1,1)_{01}\,,
\ee
where $\mathfrak{u}(1,1)_{IJ}={\rm Span}_\R\{\mathsf{i}\,R_{II}\,,\mathsf{i}\,R_{JJ}\,,\mathsf{j}\,R_{(IJ)}\,,\mathsf{k}\,R_{(IJ)}\}$.

\item In the ``massless'' case with $m=s$, 
the dual group is $\tilde G= B(\h_{\sst (1,1)},\mathbb{F})$,
and the dual coadjoint orbit is given by 
\be
  \tilde\f = (\eta_{\sst (1,1)})^{-1}\,
  \mathsf{i}\,|s|
  = 
      \mathsf{i}\,|s|\,
      \left( \cR^{00}+\cR^{11} \right)\,.
      \label{short BdS}
\ee
For $\mathbb{F}=\C$, 
the  stabiliser is
\be
     \tilde{\mathfrak{g}}_{\tilde\f}\cong
     \mathfrak{u}(1)_{R_{00}}\oplus
     \mathfrak{u}(1)_{R_{11}}\,,
\ee
while for $\mathbb{F}=\mathbb{H}$,
the stabiliser is 
\be
     \tilde{\mathfrak{g}}_{\tilde\f}\cong
     \mathfrak{u}(1,1)_{01}\,.
\ee
\end{itemize}

\item The coadjoint orbit of BdS spinning particle
with light-like momentum and space-like spin
is given by
\be
	\phi\,\O_{\sst (4)}^{\ -1}=
    (E\,\cJ^{1+}+s\,\cJ^{23})\,\O_{\sst (4)}^{\ -1}
    =\frac{1}{4}
    \begin{pmatrix}
    s\,\t_1 & -E\,(\t_1+\t_2) \\
    E(\t_1-\t_2) & s\,\t_1
    \end{pmatrix},
\ee
where $\cJ^{1+}=\frac{1}{2}(\cJ^{10}+\cJ^{14})$.
The dual group is $B(\h_{\sst (2,2)},\mathbb{F})$ and
the dual coadjoint orbit is given by
\ba
  \tilde\f \eq (\h_{\sst (2,2)})^{-1}\,
    \frac{1}{4}
    \begin{pmatrix}
      \mathsf{i}\,(|s|\,\t_2+\ve\,1_{\sst(2)}) & -\ve\,1_{\sst(2)} \\
      \ve\,1_{\sst(2)} & \mathsf{i}\,(-|s|\,\t_2+\ve\,1_{\sst(2)})
    \end{pmatrix}\nn
     \eq
        \frac{\mathsf{i}\,|s|}{2}\left(
        \cR^{zz}+\cR^{\bar z\bar z}
        -\cR^{ww}-\cR^{\bar w\bar w}\right)
        +{\mathsf{i}\,\ve}\left(
        \cR^{z\bar z}+\cR^{w\bar w}\right)\,.
        \label{BdS LS}
\ea
The dual stabiliser for $\mathbb F=\C$ is
\be
   \tilde{\mathfrak{g}}_{\tilde\f}\cong
   \mathfrak{u}(1)_{\mathsf{i}(R_{zz}+R_{\bar z\bar z})}\oplus
   \mathfrak{u}(1)_{\mathsf{i}\,R_{(z\bar z)}}\oplus
   \mathfrak{u}(1)_{\mathsf{i}(R_{ww}+R_{\bar w\bar w})}\oplus
   \mathfrak{u}(1)_{\mathsf{i}\,R_{(w\bar w)}}\,,
\ee
and the dual stabiliser for $\mathbb{F}=\mathbb H$
is generated by
\ba
   & \mathsf{i}(R_{zz}+R_{\bar z\bar z})\,,\quad 
    \mathsf{i}\,R_{(z\bar z)}\,,\quad  
   \mathsf{i}\,(R_{ww}+R_{\bar w\bar w})\,,\quad  
    \mathsf{i}\,R_{(w\bar w)}\,, \nn 
   & \mathsf{j}\,R_{(z\bar w)}\,,\quad 
    \mathsf{j}\,R_{(\bar zw)}\,,\quad 
   \mathsf{k}\,R_{(z\bar w)}\,,\quad  
    \mathsf{k}\,R_{(\bar zw)}\,.
\ea
Here, $\cV^{z/\bar z}=\frac12(\cV^{0'}\pm\mathsf{i}\,\cV^1)$ and
$\cV^{w/\bar w}=\frac12(\cV^0\pm\mathsf{i}\,\cV^2)$ 
can be viewed as lightcone basis for complex and quaternionic space: $\cR^{z\bar z}$
and $\cR^{w\bar z}$ correspond
to the nilpotent matrices $R^{z\bar z}$
and $R^{w\bar w}$, namely 
$(R^{z\bar z})^2=0$, $(R^{w\bar w})^2=0$
and $[R^{z\bar z}, R^{w\bar w}]=0$\,.
Note that $\ve\cong \l\,\ve$ for any $\l \in \R_+$, 
and its sign is determined by sign of $Es$.

\item The coadjoint orbit of BdS spinning particle
with light-like momentum and light-like spin
is given by
\be
	\phi\,\O_{\sst (4)}^{\ -1}=
    (E\,\cJ^{1+}+\e\,\cJ^{2+'})\,\O_{\sst (4)}^{\ -1}
    =\frac{1}{4}
    \begin{pmatrix}
       -E\,\t_2-\e\,\t_1 &\quad  -(E+\mathsf{i}\e)\,\t_2 \\
       -(E-\mathsf{i}\e)\,\t_2 &\quad  -E\,\t_2+\e\,\t_1
    \end{pmatrix}.
    \label{BdS LL}
\ee
The dual group is $B(\h_{\sst(1,1)},\mathbb{F})$,
and the dual coadjoint orbit is
nilpotent and given by
\be
      \tilde\f =
      ( \h_{\sst(1,1)})^{-1}\,\frac14\begin{pmatrix} 
      \mathsf{i}\,\ve & \ve \\ 
      -\ve & \mathsf{i}\,\ve
      \end{pmatrix}
	= \mathsf{i}\,\ve\,\cR^{\bar z z}\,,
\ee
with $\cV^{z/\bar z}=\frac12(\cV^0\pm \mathsf{i}\,\cV^1)$,
and $\ve\cong \l\,\ve$ for any $\l \in \R_+$.
The sign of $\ve$ is fixed by 
the sign of $E\,\e$. 
 The dual stabiliser for $\mathbb{F}=\C$ is 
\be
    \tilde{\mathfrak{g}}_{\tilde\f}\cong
    \mathfrak{u}(1)_{\mathsf{i}(R_{zz}+R_{\bar z\bar z})}\oplus
    \R_{\mathsf{i}(R_{zz}-2\,R_{\bar z z}+R_{\bar z\bar z})}\,,
\ee
and 
the dual stabiliser for $\mathbb{F}=\mathbb{H}$ is
\be
    \tilde{\mathfrak{g}}_{\tilde\f}\cong 
    \mathfrak{u}(1)_{\mathsf{i}\,R_{z\bar z}}
    \oplus\mathfrak{iso}(1,1)\,,
\ee
where $\mathfrak{iso}(1,1)$ is generated by
\be
     \mathsf{i}(R_{zz}-R_{z\bar z}+R_{\bar z\bar z})\,,\quad
     \mathsf{j}\,R_{z\bar z}\,,\quad
     \mathsf{k}R_{z\bar z}\,.
\ee

\item The coadjoint orbit of BdS particle with
space-like momentum and doubly light-like spin
is given by
\be
	\phi\,\O_{\sst (4)}^{\ -1}=
    (m\,\cJ^{12}+\e\,\cJ^{++'})\,\O_{\sst (4)}^{\ -1}=
    \frac{1}{4}
    \begin{pmatrix}
    0 & 2\,m\,\t_0 \\
    -2\,m\,\t_0 & \e\,1_{\sst(2)} +\mathsf{i}\,\e\,\t_0
    \end{pmatrix}.
\ee
Then the dual group is $B(\h_{\sst (2,2)},\mathbb{F})$,
and the dual coadjoint orbit is given by
\ba
  \tilde\f\eq 
  (  \h_{\sst(2,2)})^{-1} \, \frac{\mathsf{i}}{2}
    \begin{pmatrix}
    m &0 & 0 & 0 \\
    0 & -m+\ve & 0 & -\ve \\
    0 & 0 & -m & 0 \\
    0 & -\ve & 0 & m+\ve
    \end{pmatrix} \nn
    \eq 
     \frac{\mathsf{i}\,m}{2}
      \left( \cR^{0'0'}-\cR^{11}-4\,\cR^{(+-)}\right)
      +2\,\mathsf{i}\,\ve\,\cR^{--}\,,
      \label{BdS llS}
\ea
where $\cV^{\pm}=\frac12(\cV^0\pm \cV^2)$.
The dual stabiliser for  $\mathbb{F}=\C$ is 
\be
     \tilde{\mathfrak{g}}_{\tilde\f}\cong
     \mathfrak{u}(1)_{\mathsf{i}(R_{0'0'}-R_{11})}\oplus
     \mathfrak{u}(1)_{\mathsf{i}\,R_{(+-)}}\oplus
     \mathfrak{u}(1)_{\mathsf{i}(2R_{++}+2R_{--}+R_{(+1)}+R_{(-1)})}\oplus
     \mathfrak{heis}_2\,,
\ee

where $\mathfrak{heis}_2$ is generated by
\be
    \mathsf{i}\,(R_{0'0'}+R_{11})\,,\quad
    R_{[0'1]}\,,\quad
    \mathsf{i}\,R_{(0'1)}\,,
\ee
and the dual stabiliser for $\mathbb{F}=\mathbb{H}$ 
is generated by 
\ba
     & R_{[0'1]}\,,\quad  
     \mathsf{i}\,R_{0'0'}\,,\quad  
     \mathsf{i}\,R_{11}\,,\quad   
     \mathsf{i}\,R_{(+-)}\,,\quad  
     \mathsf{i}\,R_{(0'1)}\,,\quad    
     \mathsf{j}\,R_{(0'-)}\,,\quad  
     \mathsf{k}\,R_{(0'-)}\,,\nn
     &\mathsf{j}\,R_{(1+)}\,,\quad  
     \mathsf{k}\,R_{(1+)}\,,\quad  
     \mathsf{i}\,(2R_{++}+2R_{--}+R_{(+1)}+R_{(-1)})\,.
\ea
Note that $\ve\cong \l\,\ve$ for any $\l \in \R_+$ and its sign is given by sign of $m\,\e$.

\item The coadjoint orbit of BdS particle
with light-like momentum and doubly light-like spin
is given by
\be
	\phi\,\O_{\sst (4)}^{\ -1}=
    (E\,\cJ^{1+}+\e\,\cJ^{-+'})\,\O_{\sst (4)}^{\ -1}
    =\frac{1}{4}
    \begin{pmatrix}
      0 & -E\,(\t_2-\t_0) \\
      -E\,(\t_2+\t_0) & -\e\,(1_{\sst(2)}+\t_1)
    \end{pmatrix}.
    \label{BdS lllS}
\ee
The dual group is $B(\h_{\sst(2,1)},\mathbb{F})$,
and the dual coadjoint orbit is nilpotent and given by
\be
  \tilde\f = 
  ( \h_{\sst(2,1)})^{-1}\,\frac{\ve}{2}
  \begin{pmatrix}
    0 & -1 & 1 \\
    1 & 0 & 0 \\
    -1 & 0 & 0
  \end{pmatrix}
  =\ve\,\cR^{[0'+]}\,,
\ee
with $\cV^{\pm}=\frac12(\cV^0\pm \cV^1)$.
The dual stabiliser for $\mathbb{F}=\C$ is
\be
        \tilde{\mathfrak{g}}_{\tilde\f}\cong 
        \R_{R_{[0'-]}}\oplus
        \mathfrak{u}(1)_{\mathsf{i}\,R_{(+-)}}\oplus
        \mathfrak{u}(1)_{\mathsf{i}\,R_{--}}\,,
\ee
and the dual stabiliser for $\mathbb{F}=\mathbb{H}$
is 
\be
        \tilde{\mathfrak{g}}_{\tilde\f}\cong 
        \R_{R_{[0'-]}}\oplus
        \mathfrak{iso}(3)\,,
\ee
where $\mathfrak{iso}(3)$ is generated by imaginary generators
\be
    \mathsf{q}_\a\,R_{(+-)}\,,\quad 
    \mathsf{q}_\a\,R_{--}\,.
\ee
Note that $\ve\cong \l\,\ve$ for any $\l \in \R_{\neq 0}$.

\end{itemize}

\subsection{Particles on the conformal boundary}
The twistor group $B(\O_{\sst (4)},\mathbb F)$
is not only the isometry group of AdS$_{|\mathbb F|+3}$, but also
the conformal group in $|\mathbb F|+2$ dimensions.
The conformal particles living in 
$|\mathbb F|+2$ dimensions
have \emph{very small}
phase space compared to the other particles
living in AdS or BdS.
In this section,
we consider the twistor action for such conformal particles.

\begin{itemize}

\item The coadjoint orbit
 corresponding to conformal scalar particle
 is given by
\be 
    \phi\,\O_{\sst (4)}^{\ -1}=
    \e\,\cJ^{+'+}\,\O_{\sst (4)}^{\ -1}
     = \frac{\e}{4}
     \begin{pmatrix}
     0 & 0 & 0 & 0 \\
     0& 1 & 1 & 0 \\ 
     0& 1 & 1 & 0 \\ 
     0 & 0 & 0 & 0 
     \end{pmatrix}.
\ee
The dual group is $\tilde G= B(\h_{\sst (1)},\mathbb F)$,
and the dual coadjoint orbit is  
the trivial orbit,
\be 
    \tilde\f=0\,.
\ee

\item The coadjoint orbit of spinning singleton 
is given by
\be \label{spinning singleton1}
	\phi\,\O_{\sst (4)}^{\ -1}=
    s\,(
    \cJ^{0'0}
    \pm \cJ^{12}
    +\cJ^{34}\mp \cJ^{56})\,\O_{\sst (4)}^{\ -1}
    =\frac{s}{2}
    \begin{pmatrix}
    1_{\sst(2)} + \mathsf{i}\,\t_0 && \mp(\mathsf{i}\,1_{\sst(2)}-\t_0) \\
    \pm(\mathsf{i}\,1_{\sst(2)} -\t_0)  && 1_{\sst(2)}+\mathsf{i}\,\t_0
    \end{pmatrix}.
\ee
The dual group is $\tilde G=
B(\h_{\sst (1)},\mathbb F)$,
and the dual coadjoint orbit is a single point,
\be  \label{singleton dual rep1}
    \tilde\f=\pm 2\,|s|\,\mathsf{i}\,.
\ee
Note here that we use  complex entries  
for the coadjoint element $\phi$, and hence
the corresponding coadjoint orbit does not exist for $\mathbb F=\R$,
as is also manifest from the use of the 
elements $\cJ^{34}$ and $\cJ^{56}$.

For $\mathbb F=\C$,
each sign choices correspond
to distinct cases:
they are the massless helicity $\pm s$ particles. 
Remark also that we included
the $\mathfrak{u}(1)$ part
of $\mathfrak{b}(\O_{\sst (4)},\mathbb C)\cong \mathfrak{u}(2,2)$
through $\cJ^{56}$
in a way that the dual group becomes minimal,
that is $B(\h_{\sst (1)},\mathbb C)
\cong U(1)$,
but 
other choices of $\mathfrak{u}(1)$ are equally viable with the dual group 
$B(\h_{\sst (1,3)},\mathbb C)\cong U(1,3)$.

For $\mathbb F=\mathbb H$,
the two sign choices are in fact equivalent:
they are related by a $\pi$ rotation 
by, for instance, $J^{15}$, which is a 
quaternionic matrix.
Then, the opposite helicity can be described
by the same action but with
a different isomorphism between $B(\O_{\sst (4)},\mathbb H)=O^*(8)$ and $\widetilde{SO}^+(2,6)$.

\end{itemize}

\section{Classification
of twistor particles according 
to the dual groups}
\label{sec:summary}
Let us summarize the results of classification
according to the dual group $\tilde G=B(\eta_{\sst (p,M-p)},\mathbb F)$.
Since the twistor group is 
$G=B(\O_{\sst (4)},\mathbb F)$,
the possible values of $M$ are
restricted to 1, 2, 3 and 4.

\subsection{Compact dual group}
All the ordinary twistor particles, namely the massive and massless, twistor particles in AdS$_{|\mathbb F|+3}$, 
as well as the conformal twistor particles in 
$|\mathbb F|+2$ dimensions, 
are obtained with compact dual groups 
$\tilde G=B(\eta_{\sst (M)},\mathbb F)$.

Even though we have limited our focus
to the symmetric spin case,
we can fully generalize
the analysis 
when the dual group is compact.
This is thanks to the mileage that 
we have in
the representation theory side:
we know all the unitary irreducible representations associated with
the dual pair correspondence 
between $B(\O_{\sst (4)}, \mathbb F)$
and $B(\eta_{\sst (M)},\mathbb F)$ \cite{Basile:2020gqi} 
(also \cite{Kashiwara1978, Howe1989i, Howe1989ii}).
Let us summarize the results  here:
the correspondences read
for $\mathbb F=\mathbb R$,
\ba 
    (\tfrac{\bar n+1}2;\tfrac{\bar n}2)_{Sp(4,\mathbb R)}\ &\leftrightarrow & 
    \  
    [\bar n]_{O(1)}\,,
    \label{O(1)} \\
    (s+1;s)_{Sp(4,\mathbb R)}\ &\leftrightarrow & 
    \
    [2s]_{O(2)}\,,\\
   (s+\tfrac32;s)_{Sp(4,\mathbb R)}\ &\leftrightarrow & 
    \
    [2s]_{O(3)}\,,\\ 
     (s+\tfrac52;s)_{Sp(4,\mathbb R)}\ &\leftrightarrow &
    \
    [2s+1,1]_{O(3)}\,,\\ 
     (s+n+2;s)_{Sp(4,\mathbb R)}\ &\leftrightarrow & 
    \
    [2s+n,n]_{O(4)}\,,
\ea
and for $\mathbb F=\mathbb C$, 
\ba 
    (s+1;s,\pm s)_{SU(2,2)}\ &\leftrightarrow & 
    \
    [\pm 2s]_{U(1)}\,,\\
     (s_1+2;s_1,s_2)_{SU(2,2)}\ &\leftrightarrow & 
    \
    [s_1+s_2,s_2-s_1]_{U(2)}\,,
     \label{U2}\\
   (s+n+2;s,\pm s)_{SU(2,2)}\ &\leftrightarrow & 
    \
    [\pm(2s+n),\pm n]_{U(2)}\,,
    \label{U2'}\\
     (s_1+n+3;s_1,s_2)_{SU(2,2)}\ &\leftrightarrow & 
    \ 
    \left\{
    \begin{array}{c}
    \,[s_1+s_2+n,n,s_2-s_1]_{U(3)} 
   \\
    \,[s_1+s_2,-n,s_2-s_1-n]_{U(3)} 
    \end{array}
    \right.
    ,\label{U3}\\
     (s_1+n+k+4;s_1,s_2)_{SU(2,2)}\ &\leftrightarrow & 
    \
    [s_1+s_2+n,n,-k,s_2-s_1-k]_{U(4)}\,,
    \label{U4}
\ea
and finally for $\mathbb F=\mathbb H$,
\ba 
    (s+2;s,s,\pm s)_{O^*(8)}\ &\leftrightarrow & 
    \
    [2s]_{Sp(1)}\,,\\
     (s_1+4;s_1,s_2,\pm s_2)_{O^*(8)}\ &\leftrightarrow & 
    \
    [s_1+s_2,s_1-s_2]_{Sp(2)}\,,
    \label{Sp2}\\ 
   (s_1+s_2-s_3+6;s_1,s_2,\pm s_3)_{O^*(8)}\ &\leftrightarrow & 
    \
    [s_1+s_2,s_1-s_3,s_2-s_3]_{Sp(3)}\,,
    \label{Sp3}\\
     (s_1+s_2-s_3+8+2n;s_1,s_2,\pm s_3)_{O^*(8)} \hspace{-20pt} &&\nn
    && \hspace{-80pt} \leftrightarrow\ 
[s_1+s_2+s_3+n,s_1+n,s_2+n,s_3+n]_{Sp(4)}\,.\label{Sp4}
\ea
In the last case of $\mathbb F=\mathbb H$, 
the $\pm$ sign corresponds
to two possible isomorphisms between $O^*(8)$
and $\widetilde{SO}^+(2,6)$.
The label of $B(\eta_{\sst (4)},\mathbb{F})$ is $(\Delta; s_1,\ldots)$
where $\Delta$ is the minimum of the global energy in AdS or equivalently 
the conformal weight,
and $(s_1,\ldots)$ is the Young diagram representing the compact rotational subgroup $\widetilde{SO}(|\mathbb F|+2)$.
The label $[\ell_1,\ldots]$
is the Young diagram  representing the compact 
dual group $ B(\eta_{\sst (M)},\mathbb F)$, that is,
the eigenvalues of the Cartan subalgebra generators for the highest weight state.
From these results, we can extract  the information
about their classical counterpart.

Below, we summarize the classifications we found in the previous section,
which cover only symmetric spin cases.
Then, we complete the classifications by making use of the results
\eqref{O(1)}--\eqref{Sp4}
obtained in the dual pair correspondence.

\begin{itemize}

\item $B(\eta_{\sst (1)},\mathbb F)$ \\
For $M=1$, we find conformal twistor particles in 
$|\mathbb F|+2$ dimensions
with the action,
\be
    S=\int \left[ \bar\p_\a\,\rd\x^\a-\bar\x^{\dot\a}\,\rd\p_{\dot\a} 
    +A\left(\bar\p_\a\,\x^\a-\bar\x^{\dot\a}\,\p_{\dot\a} - 2\,\mathsf{i}\,s \right)\right]_\R\,,
\ee
which reproduces
earlier results 
\cite{Ferber:1977qx, Shirafuji:1983zd, Bengtsson:1987si,Adams1987, Townsend:1991sj, Howe:1992bv}.
The constraints are associated with the moment maps
generating $\hat{\mathbb F}\cong \mathfrak{b}(\eta_{\sst (1)},\mathbb F)
$ symmetry.

For $\mathbb F=\R$, these constraints
disappear, and the system describes 
the conformal scalar in 3 dimensions.
In fact, the same particle action describes
also the conformal spinor in 3 dimensions. The
distinction between the conformal scalar and spinor can be also made based on the dual group $O(1)=\mathbb Z_2$. Since the latter is a finite group,
the classical particle action itself cannot distinguish the two, but when we perform quantization,
the $\mathbb Z_2$ symmetry can be imposed as constraints. The even and odd parts will result in the conformal scalar and the conformal spinor, respectively. 

For $\mathbb F=\C$, the action describes 4d conformal particles with helicity $s$: positive and negative signs of $s$
correspond to positive and negative helicity, respectively.

For $\mathbb F=\mathbb H$, it describes 6d conformal
anti-chiral particles with spin $(s,-s)$. 
In this case, the two signs of $s$ are equivalent and
describe the same particle.
For the description of a chiral particle
with spin $(s,s)$,
we can use the same action, but
different isomorphism between
$\widetilde{SO}^+(2,6)$ 
and $O^*(8)$.

The dual coadjoint orbit is trivial
for $\mathbb F=\mathbb R$ as the dual coadjoint space itself is trivial.
For $\mathbb F=\mathbb C$, the dual orbit 
consists of a single point.
For $\mathbb F=\mathbb H$,
it is $S^2\cong Sp(1)/U(1)$\,.

\item  $B(\eta_{\sst (2)},\mathbb F)$  \\
For $M=2$, we find the massless spin $s$ particles
in AdS$_{|\mathbb F|+3}$, with the action,
\ba
    S\eq \int \,\Big[ \,
    \bar\p_{\a\,1}\,\rd\x^\a{}_1-\bar\x^{\dot\a}{}_1\,\rd\p_{\dot\a}{}_1
    +A^{11}\left(\bar\p_\a{}_1\,\x^\a{}_1-\bar\x^{\dot\a}{}_1\,\p_{\dot\a}{}_1 \right)
    \nn 
    &&\qquad +\, \bar\p_{\a\,2}\,\rd\x^\a{}_2-\bar\x^{\dot\a}{}_2\,\rd\p_{\dot\a}{}_2
    +A^{22}\left(\bar\p_\a{}_2\,\x^\a{}_2-\bar\x^{\dot\a}{}_2\,\p_{\dot\a}{}_2 \right)
    \nn 
     && \qquad 
    +\,A^{(12)}\left(\bar\p_\a{}_{(1}\,\x^\a{}_{2)}-\bar\x^{\dot\a}{}_{(1}\,\p_{\dot\a}{}_{\,2)}
    \right)\nn
    && \qquad 
    +\,A^{[12]}
    \left(\bar\p_\a{}_{[1}\,\x^\a{}_{\,2]}-\bar\x^{\dot\a}{}_{[1}\,\p_{\dot\a}{}_{\,2]}
    -s \right)\Big]_\R. 
    \label{massless twistor}
\ea
Each of the constraints $\tilde\mu_{11}$, $\tilde \mu_{22}$ and $\tilde \mu_{(12)}$ 
generates $\hat{\mathbb F}$,
whereas the moment map $\tilde \mu_{[12]}$ generates
$U(1)\cong O(2)$
(remind $\tilde\mu_{IJ}
=\bar\pi_{\alpha\,I}\, 
	\xi^\alpha{}_J-\bar\xi^{\dot\alpha}{}_I\,\pi_{\dot\alpha\,J}$).
This action
reproduces the earlier results 
of AdS$_{|\mathbb F|+3}$
massless particles
\cite{Arvanitakis:2016wdn,
Arvanitakis:2016vnp,
Arvanitakis:2017cpk},
where the difference is that the non-zero spin is introduced by SUSY, instead of shifting a constraint by spin label $s$.

For $\mathbb F=\R$, the 
constraints associated with
$\tilde\mu_{11}$, $\tilde \mu_{22}$ and $\tilde \mu_{(12)}$ are absent,
whereas  the $U(1)$ constraint
$\tilde\mu_{[12]}\approx s$  fixes
the spin.
The dual coadjoint orbit is simply
a single point in this case.

For $\mathbb F=\C$, the spin is fixed by 
the $U(1)$ of the $SU(2)$ generated by
$\tilde \mu_{11}-\tilde \mu_{22}$ and
$\tilde\mu_{12}$,
defining a coadjoint orbit
$SU(2)/U(1)\cong S^2$,
whereas the diagonal $U(1)$ generated by $\tilde \mu_{11}+\tilde \mu_{22}$ has a trivial coadjoint orbit, that is, the constraint
is given without any constant shift.

For $\mathbb F=\mathbb H$,
the dual group $B(\eta_{\sst (2)},\mathbb H)$ is $Sp(2)$,
and the constraints 
define a six dimensional coadjoint orbit
$Sp(2)/[U(1)\times Sp(1)]$.

In fact, for $\mathbb F=\C$
and $\mathbb H$,
any system of constraints
can be mapped to
\be 
    A^{11}\,(\tilde\mu_{11}-\mathsf{i}\,\ell_1)
    +A^{22}\,(\tilde\mu_{22}-\mathsf{i}\,\ell_2)
    +A^{(12)}\,\tilde\mu_{(12)}
    +A^{[12]}\,\tilde\mu_{[12]}\,,
    \label{mixed sym constraints}
\ee
with $\ell_1\ge \ell_2$,
using a suitable transformation of 
$B(\eta_{\sst(2)},\mathbb F)$,
that is, $U(2)$
or $Sp(2)$.
Note that 
for $\mathbb F=\mathbb H$
we can change the sign of $\ell_i$ 
by, for instance, an adjoint action of $\mathsf{j}$ on the $ii$ component,
and hence 
positive and negative $\ell_i$ are equivalent: 
both signs correspond
to the same orbit.
On the contrary,
for $\mathbb F=\mathbb C$,
positive and negative $\ell_i$
correspond to distinct orbits.

The massless spin $s$ action \eqref{massless twistor} 
corresponds to the case with $[\ell_1,\ell_2]=[s,-s]$,
while we may consider the more general case with $[\ell_1, \ell_2]$.

Let us first consider the case with $\ell_2\le 0$.
According to \eqref{U2} and \eqref{Sp2},
the action with the constraints 
\eqref{mixed sym constraints}
describe massless particle of mixed symmetry spin $(s_1, s_2)$ in AdS$_5$
and 
$(s_1,s_2, \pm s_2)$ in AdS$_7$,
for $s_1>|s_2|$.
The two signs reflect the two possible isomorphisms beween $O^*(8)$ and $\widetilde{SO}^+(2,6)$.
The labels 
$(s_1, s_2)$ and $[\ell_1, \ell_2]$ are related by
\be 
    [\ell_1,\ell_2]=[s_1+s_2, s_2-s_1]
    =s_1\,[1,-1]+s_2\,[1,1]\,.
    \label{massless}
\ee 
Remark that for $s_1=|s_2|$ 
the action with \eqref{massless}
describes in fact a massive particle:
the quantization of the model would generate
a small finite mass.

Let us now turn to the remaining case
of $\ell_2>0$.
For $\mathbb F=\mathbb H$,
a positive $\ell_2$ 
is equivalent to a negative $\ell_2$, as we 
explained above.
Therefore, it is sufficient to consider
the $\mathbb F=\mathbb C$  case only.
According to \eqref{U2'}, this new case corresponds to massive particle of 
mass $m>s$ and spin $(s,\pm s)$,
where 
$[\ell_1, \ell_2]$
is parameterized by
\be 
    [\ell_1,\ell_2]=
  [\pm m+s, \pm m-s]
    =s\,[1,-1] \pm m\,[1,1]\,.
    \label{massive}
\ee 
Remind that in the case of massless 
particle of symmetric spin $(s,0)$ \eqref{massless twistor}, the $U(1)$ part of the constraints does not have a constant shift. 
In the case of massless particle
of mixed symmetry spin $(s_1,s_2)$
\eqref{massless}
or massive particle of mixed symmetry spin $(s,\pm s)$ 
\eqref{massive},
the $U(1)$ part has 
the shift $2s_2$ or $\pm 2m$.
In other words, 
the first spin label $s_1$ is always fixed by
the $SU(2)$ orbit data $s$,
whereas the $U(1)$ orbit position,
say $n$, determines
the second spin label as $s_2=\frac{|n|}2$ for 
$\frac{|n|}2< s$.
For $\frac{|n|}2\ge s$, 
it rather determines the mass value as 
$m=\frac{|n|}2$
while the second spin label is fixed
by $s_2={\rm sgn}(n)\,s$.
The latter type of massive twistor action 
in AdS$_5$ was considered in  
\cite{Claus:1999zh} for the scalar case with $s=0$,
and its quantization was discussed in \cite{Claus:1999jj}.
See e.g. \cite{Arvanitakis:2017cpk} for an early account 
for the AdS$_5$ exception.
Let us remark that
the distinction between the two cases
$|n|\ge 2s$ and $|n|<2s$
may invoke subtle issues when
spin is introduced by SUSY,
and this point has not been addressed in earlier literature.

\item $B(\eta_{\sst (3)},\mathbb F)$ \\
For $M=3$, we find 
the case of entangled mass and spin
\eqref{entangled} with $s\,\epsilon >0$.
The action is given simply by
\ba
    S\eq \int \,\Big[ \,
    \delta^{IJ}\left(
    \bar\p_{\a\,I}\,\rd\x^\a{}_J-\bar\x^{\dot\a}{}_J\,\rd\p_{\dot\a}{}_I\right)
    +A^{(IJ)}\,\tilde\mu_{(IJ)}
       \nn 
    &&\qquad +\, 
      A^{[12]}\left(\tilde \mu_{[12]}-s\right)+
   A^{[23]}\,\tilde\mu_{[23]}
    +A^{[31]}\,\tilde\mu_{[31]}
    \Big]_{\mathbb R}\,,
    \label{M=3}
\ea
where $I, J=1,2,3$.
This action  describes 
the very light massive particle
of spin $s$: the
corresponding orbit of AdS$_{|\mathbb F|+3}$ is 
the massive part of the massless limit of the massive spinning orbit,
which arises together with the massless orbit.
Let us refer to this special orbit
as the orbit of ``very light mass''.
The very light massive orbits
have the same classical Casimir values
as the massless ones,
so we cannot distinguish 
the very light massive particles 
from the massless ones in terms of the classical mass parameter.
If we quantize the particle actions,
the classical mass parameters will acquire 
a small quantum correction, from which we can distinguish the two cases.

For $\mathbb F=\C$ and $\mathbb H$,
we can perform a $B(\eta_{\sst (3)},
\mathbb F)$ rotation to transform
the constraints in a way that
the part with non-trivial shifts are
\be  
    A^{11}\,(\tilde\mu_{11}-\mathsf{i}\,\ell_1)
    +A^{22}\,(\tilde\mu_{22}-\mathsf{i}\,\ell_2)
     +A^{33}\,(\tilde\mu_{33}-\mathsf{i}\,\ell_3)\,,
\ee 
with $\ell_1\ge \ell_2\ge \ell_3$.
Again, for $\mathbb F=\mathbb H$, the sign of $\ell_i$ can be set to be 
non-negative.
Note that the case \eqref{M=3}
corresponds to $[\ell_1,\ell_2,\ell_3]=[s,0,-s]$\,,
while a more general case  
$[\ell_1,\ell_2,\ell_3]=[s_1+s_2,-s_2,-s_1]$
corresponds to the light massive particle of mixed symmetry spin $(s_1,s_2)$,
according to the first case of \eqref{U3}.
For $\mathbb F=\mathbb H$,
we can turn on even the third spin label $s_3$,
 according to \eqref{Sp3},
to describe the particles of mixed symmetry spin $(s_1,s_2,s_3)$ by
the action with the constraints,
\be
    [\ell_1,\ell_2,\ell_3]=[s_1+s_2,s_1-s_3,s_2-s_3]\,.
\ee 
For $\mathbb F=\C$,
the  label $m\ge s_1$ in
\be
    [\ell_1,\ell_2,\ell_3]=\left\{
    \begin{array}{c}
    \,[m+s_2,\ m-s_1,\ s_2-s_1]
    \vspace{5pt} \\
     \,[s_1+s_2,\ -m+s_1,\ s_2-m]
    \end{array}
    \right..
\ee 
corresponds to the mass of AdS$_5$ particle of
mixed symmetry spin $(s_1,s_2)$,
according to \eqref{U3}.
Note that
the two options for $[\ell_1,\ell_2,\ell_3]$
describe exactly the same particle,
and they can be distinguished only by the $U(1)$ shift value $2(s_2\mp (s_1 - m))$.

\item  $B(\eta_{\sst (4)},\mathbb F)$ 

For $M=4$, we find 
the action for a particle of 
mass $m$ and spin $s$ as
\ba
    S\eq \int \,\Big[ \,
    \delta^{IJ}\left(
    \bar\p_{\a\,I}\,\rd\x^\a{}_J-\bar\x^{\dot\a}{}_J\,\rd\p_{\dot\a}{}_I\right)+A^{(IJ)}\,\tilde
 \mu_{(IJ)} 
    \nn
    &&\qquad +\, A^{[12]}\left(\tilde \mu_{[12]}-\frac{|m+s|}2\right)
    +A^{[34]}\left(\tilde\mu_{[34]}
    -\frac{|m-s|}2\right)
   \nn 
    &&\qquad +\, 
    A^{[13]}\,\tilde\mu_{[13]}
    +A^{[14]}\,\tilde\mu_{[14]}
    +A^{[23]}\,\tilde\mu_{[23]}
    +A^{[24]}\,\tilde\mu_{[24]}
    \Big]_{\mathbb R}\,,
    \label{M=4}
\ea
where $I, J=1,2,3, 4$.
For $s=0$, this action
reproduces
the scalar twistor model 
\cite{Cederwall:2000km},
where the dual group symmetry
was referred to as R-symmetry.

Again, for $\mathbb F=\C$ and $\mathbb H$,
with a $B(\eta_{\sst (4)},
\mathbb F)$ rotation,
the action can be transformed
so that the constraints with non-trivial shifts are
\be  
    A^{11}\,(\tilde\mu_{11}-\mathsf{i}\,\ell_1)
    +A^{22}\,(\tilde\mu_{22}-\mathsf{i}\,\ell_2)
     +A^{33}\,(\tilde\mu_{33}-\mathsf{i}\,\ell_3)
     +A^{44}\,(\tilde\mu_{44}-\mathsf{i}\,\ell_4)\,,
\ee 
where the case \eqref{M=4}
corresponds to $[\ell_1,\ell_2,\ell_3,\ell_4]
    =[\frac{s+m}2,\frac{-s+m}2,-\frac{-s+m}2,-\frac{s+m}2]$.
For $\mathbb F=\mathbb C$, this can 
be generalized to
\be
    [\ell_1,\ell_2,\ell_3,\ell_4]
    =\left[\tfrac{s_1+2s_2+m+n}2,
    \tfrac{-s_1+m+n}2,-\tfrac{-s_1+m-n}2,-\tfrac{s_1-2s_2+m-n}2\right],
\ee 
and the corresponding action describes
a AdS$_5$ particle of mass $m$ and mixed symmetry
spin $(s_1,s_2)$, according to \eqref{U4}.
Note that $n$ only enters the $U(1)$ 
constraint as the shift $2(s_2+n)$.
For $\mathbb F=\mathbb H$, the general case,
\be
    [\ell_1,\ell_2,\ell_3,\ell_4]
    =\left[\tfrac{s_1+s_2+s_3+m}2,
    \tfrac{s_1-s_2-s_3+m}2,\tfrac{s_2-s_3-s_1+m}2,\tfrac{s_3-s_1-s_2+m}2\right],
\ee
describes a AdS$_7$ particle of 
mass $m$ and mixed symmetry spin $(s_1,s_2,\pm s_3)$, according to \eqref{Sp4}.
Again, the $\pm$ signs correspond
to two isomorphisms between $O^*(8)$ and $\widetilde{SO}^+(2,6)$.

\end{itemize}

\subsection{Non-compact dual group}
In the case of non-compact dual groups,
we find various exotic particle species.
In this case,
we have very limited information about
unitary irreducible representations
because 
they
are not of highest weight type
and
the relevant dual pair correspondence is much harder to analyze.
We can regard the situation 
as the mileage of the classical analysis
for the representation theory.
Let us remind again that we have limited to the case with two labels for simplicity, 
but the full cases can be treated by the same method.

\begin{itemize}
    
\item $B(\eta_{\sst (1,1)},\mathbb F)$

In this case, we find 
short ``massless'' spinning tachyon
\eqref{short tachyon}.
This is to be compared with
the massless spinning particles with $B(\eta_{\sst (2)},\mathbb F)$,
which is the shortening point of massive spinning particle.
For $\mathbb F=\mathbb C$ and $\mathbb H$,
we also have BdS particles:
the short ``massless'' one with space-like spin \eqref{short BdS}
and the one
with light-like momentum and spin
\eqref{BdS LL}.
Note that for $\mathbb F=\mathbb C$
with $B(\eta_{\sst (1,1)},\mathbb C)=U(1,1)$,
the three cases 
\eqref{short tachyon}, 
\eqref{short BdS}, and \eqref{BdS LL}
correspond respectively
to the two-sheet hyperboloid,
one-sheet hyperholoid, and cone
coadjoint orbits of $SU(1,1)$.

\item $B(\eta_{\sst (2,1)},\mathbb F)$

This case includes
the maximal nilpotent orbit,
namely
the BdS particle of light-like momentum
and doubly-light-like spin
\eqref{BdS lllS}.
Besides it, we 
find also large remnant orbits
\eqref{entangleSN} and \eqref{entangleTN}
appearing in the short ``massless'' limit
of the BdS particle (the $|m|<|s|$ case 
of massive spinning partlce) and the tachyon \eqref{mu nu tachyon}.
The latter two cases are to be compared with
the case of 
$B(\eta_{\sst (3)},\mathbb F)$, namely 
 the large remnant orbit
appearing in the massless limit of massive spinning orbit.

\item $B(\eta_{\sst (2,2)},\mathbb F)$

This case 
includes
all kinds of ``long'' tachyons:
the ones with time-like spin \eqref{tachyonT},
 space-like spin \eqref{tachyonS},
and light-like spin \eqref{tachyonN}.
Moreover,
it contains
all kinds of ``long'' BdS particles:
the one appearing as the $|m|<|s|$ subcase
of massive spinning particle,
the ones with 
space-like momentum and space-like spin \eqref{BdS},
light-like momentum and space-like spin
\eqref{BdS LS}
and 
space-like momentum and doubly-light-like spin
\eqref{BdS llS}.

\item $B(\eta_{\sst (3,1)},\mathbb F)$

Finally,
this case contains 
the massive, massless and tachyonic
continuous spin particles \eqref{conti}.
The action reads
\ba
    S\eq \int \,\Big[ \,
    \eta^{IJ}\left(
    \bar\p_{\a\,I}\,\rd\x^\a{}_J-\bar\x^{\dot\a}{}_J\,\rd\p_{\dot\a}{}_I\right)+A^{(IJ)}\,\tilde
 \mu_{(IJ)} 
    \nn
    &&\qquad +\, A^{[01]}\left(\tilde \mu_{[01]}-
    |\nu|\right)
    -A^{[23]}\left(\tilde\mu_{[23]}
    -|s|\right)
   \nn 
    &&\qquad +\, 
    A^{[02]}\,\tilde\mu_{[02]}
    +A^{[03]}\,\tilde\mu_{[03]}
    +A^{[12]}\,\tilde\mu_{[12]}
    +A^{[13]}\,\tilde\mu_{[13]}
    \Big]_{\mathbb R}\,,
    \label{M=3,1}
\ea
where $I,J=0,1,2,3$ and $\eta^{IJ}={\rm diag}(-1,+1,+1,+1)$.
It can be viewed as 
the twistorial counterpart
of the vectorial description
of the continuous spin particles \cite{Basile:2023vyg},
which are related to a variety of continuous spin fields identified by R. Metsaev
\cite{Metsaev:2016lhs}.
See also \cite{Buchbinder:2024hea}
for other vectorial models of (A)dS$_4$ continuous spin particles.

\end{itemize}

\section{Dimensional reduction}
\label{sec:restriction}

Many of the twistor literatures 
concern about massless and massive particles in flat spacetime,
Mink$_{|\mathbb F|+2}$.
In our construction, only conformal
particles are defined in
$|\mathbb F|+2$ dimensions,
namely
Mink$_{|\mathbb F|+2}$, AdS$_{|\mathbb F|+2}$
and dS$_{|\mathbb F|+2}$
boundaries
of AdS$_{|\mathbb F|+3}$.
These conformal particles are
massless but their spins are 
rather special carrying the  $(|\mathbb F|-1)$-row box Young diagram representation.
In fact, more general type of particles in
Mink$_{|\mathbb F|+2}$, AdS$_{|\mathbb F|+2}$
and dS$_{|\mathbb F|+2}$ can be obtained 
from AdS$_{|\mathbb F|+3}$ particles
by restricting the isometry $B(\O_{\sst (4)},\mathbb F)\cong \widetilde{SO}^+(2,|\mathbb F|+2)$
to the subgroups $H_\L$\,, that is,
$H_{\L=0}\cong \widetilde{ISO}^+(1,|\mathbb F|+1)$,
$H_{\L<0}\cong\widetilde{SO}^+(2,|\mathbb F|+1)$,
and $H_{\L>0}\cong \widetilde{SO}^+(1,|\mathbb F|+2)$.
Remark that
the conformal particles,
namely singletons, are precisely the ones
which stay irreducible
under this restriction.
The restriction of $G=B(\O_{\sst (4)},\mathbb F)$ to the subgroup $H_\L$ defines a new dual group $\tilde H_\L$ which include
the original dual group $\tilde G=B(\eta_{(q,M-q)},\mathbb F)$.
This situation can be depicted by
the so-called see-saw diagram:
\begin{equation}
    \parbox{200pt}{
    \begin{tikzpicture}
    \draw [<->] (0,0) -- (1,0.8);
    \draw [<->] (0,0.8) -- (1,0);
    \node at (-1.5,1.2) {$G=B(\O_{\sst (4)},\mathbb F)$};
    \node at (-1.5,0.4) {$\cup$};
    \node  at (-1.5,-0.4) {$H_\L$};
    \node at (2.5,-0.4) {$\tilde G=B(\eta_{\sst (q,M-q)},\mathbb F)$};
    \node at (2.5,0.4) {$\cup$};
    \node at (2.5,1.2) {$\tilde H_{\L}$};
    \end{tikzpicture}}.
    \label{restriction}
\end{equation}
The see-saw duality asserts that
the restriction of unitary irreducible representations on the left hand side
is governed by the restriction 
on the right hand side, and vice versa. More precisely,
for the irreps $\pi^G_\l$ of $G$ 
and $\pi^{\tilde H_\L}_{\t}$ of $\tilde H_\L$ 
branching respectively to $H_\L$ 
and  $\tilde G$ as
\be 
    \pi^G_\l=\bigoplus_{\s} \cN^\s_\l\,\pi^{H_\L}_\s\,,
    \qquad 
    \pi^{\tilde H_\L}_\t=\bigoplus_{\r} \tilde\cN^\r_\t\,\pi^{\tilde G}_\r\,,
\ee 
the multiplicities $\cN^\s_\l$
and $\tilde\cN^\r_\t$ are relate by 
\be 
    \cN^\s_\l
    =\tilde \cN^{\theta(\l)}_{\phi(\s)}\,,
\ee 
where $\theta$ and $\phi$
are the bijective maps of the $(G,\tilde G)$ irreps and the $(H_\L,\tilde H_\L)$ irreps, respectively.
Here, we consider the classical counterpart of the see-saw dual pair, which we may refer to as
\emph{symplectic see-saw reduction}.
The irreps correspond to coadjoint orbits,
and the restriction of an irrep down to
the irreps of a subgroup corresponds
to the foliation of a coadjoint orbits into 
the coadjoint orbits of the subgroup.
Remark that the restriction to $H_{\L=0}$ 
to obtain massive twistor particles has been discussed already in \cite{Arvanitakis:2017cpk}. Here,
we generalize the idea with extra emphasizes on see-saw dual pairs.

Let us consider the reductive subgroup
$GL(2,\mathbb F)$ of $B(\O_{\sst (4)},\mathbb F)$, 
\be
    L^\a{}_\b=\xi^\a{}^{I}\,\bar\pi_{\b\,I}\,,
	\qquad 
	\bar L_{\dot\a}{}^{\dot\b}=\pi_{\dot\a}{}^{I}\,\bar\xi^{\dot\b}{}_{I}\,.
\ee 
We may separate each of $L^\a{}_\b$ and $\bar L_{\dot\a}{}^{\dot\b}$
into the traceless and trace part:
the traceless part can be expressed
as
\be
	L_{(\a\b)}=\xi_{(\a}{}^{I}\,\bar\pi_{\b)\,I}\,,
	\qquad 
	\bar L_{(\dot\a\dot\b)}=\pi_{(\dot\a}{}^{I}\,\bar\xi_{\dot\b)I}\,,
\ee
by lowering indices with the Levi-Civita symbols,
which are Lorentz invariant tensors
only for $\mathbb F=\R$ and $\mathbb C$.
One combination of the traces is 
the center of $GL(2,\mathbb F)$,
\be 
    D=L^{\a}{}_\a+\bar L_{\dot\a}{}^{\dot\a}
    =\xi^{\a\,I}\,\bar\pi_{\a\,I}
    +\pi_{\dot\a}{}^{I}\,\bar \xi^{\dot\b}{}_{I}\,,
\ee 
which is nothing but the dilatation generator.
The other combination of the traces,
\be 
    E=L^{\a}{}_\a-\bar L_{\dot\a}{}^{\dot\a}
    =\xi^{\a\,I}\,\bar\pi_{\a\,I}
    -\pi_{\dot\a}{}^{I}\,\bar \xi^{\dot\b}{}_{I}\,,
    \label{E quart}
\ee 
takes value in $\hat{\mathbb F}$:
it vanishes for $\mathbb F=\R$, generates
the central $U(1)$ for $\mathbb F=\mathbb C$.
For $\mathbb F=\mathbb H$, the three real generators in $E$
are parts of the Lorentz subgroup $SL(2,\mathbb H)$.
Since we consider the $|\mathbb F|+2$ dimensional isometry subgroup of $B(\O_{\sst (4)},\mathbb F)$,
the dilatation generator $D$ will be discarded from the $\mathfrak{h}_\L$.
Moreover, among the translation and special conformal generators,
\be 
P^{\a\dot\b}=\xi^{\a\,I}\,\bar\xi^{\dot\b}{}_{I}\,,
\qquad 
K_{\dot\a\b}=\pi_{\dot\a}{}^{I}\,\bar\pi_{\b\,I}
\ee 
we retain only the linear combinations
\be 
    P^{\a\dot\b}+\L\,K^{\dot\b\a}\,.
\ee 
As we consider a subgroup $H_\L$,
the dual group enhances from $\tilde G$ to
$\tilde H_\L$\,. This means
that on top of the $\tilde G$ generators,
\begin{equation}
\tilde\mu_{IJ}=\bar\pi_{\alpha\,I}\, 
	\xi^\alpha{}_J-\bar\xi^{\dot\alpha}{}_I\,\pi_{\dot\alpha\,J}\,,
\end{equation}
we have additional generators that commute
with $H_\L$.
It is not difficult to enlist all possible candidates because the invariance
under $L_{(\a\b)}$ and $\bar L_{(\dot\a\dot\b)}$ implies that
the $\a,\b$ indices should be contracted:
the possible moment maps are
\be 
    \x^\a{}_I\,\x_{\a\,J}\,,
    \qquad 
    \bar\pi_{\a\,I}\,\xi^\a{}_J\,,
    \qquad 
    \bar\pi^{\a}{}_I\,\bar\pi_{\a\,J}\,,
    \label{IJ ansatz}
\ee 
and their $*$ conjugates.
For $\mathbb F=\R$ and $\mathbb C$, we can proceed to check the translational invariance to find that the additional dual generators are
\be 
    M_{IJ}=\x^\a{}_I\,\x_{\a\,J}
        -\L\,\bar\pi^{\a}{}_I
        \,\bar\pi_{\a\,J}\,,
        \qquad 
     \bar M_{IJ}=\bar\x^{\dot\a}{}_I\,\bar\x_{\dot\a\,J}
        -\L\,\pi^{\dot\a}{}_I
        \,\pi_{\dot\a\,J}\,.
\ee 
For $\mathbb F=\R$,
the original dual generators form $O(p,M-p)$,
and real $M_{IJ}$ are antisymmetric
hence span a $M(M-1)/2$ dimensional space.
The whole dual group becomes a real form of $O_{M}\times O_M$. For the $p=0$ example, we find
the sea-saw pairs for $\L<0$ and $\L>0$ as
\begin{equation}
    \parbox{140pt}{
    \begin{tikzpicture}
    \draw [<->] (0,0) -- (1,0.8);
    \draw [<->] (0,0.8) -- (1,0);
    \node at (-1,1.2) {$Sp(4,\mathbb R)$};
    \node at (-1,0.4) {$\cup$};
    \node  at (-1,-0.4) {$Sp(2,\mathbb R)^2$};
    \node at (2,-0.4) {$O(M)$};
    \node at (2,0.4) {$\cup$};
    \node at (2,1.2) {$O(M)^2$};
    \end{tikzpicture}
    },
    \qquad \quad 
    \parbox{130pt}{
    \begin{tikzpicture}
    \draw [<->] (0,0) -- (1,0.8);
    \draw [<->] (0,0.8) -- (1,0);
    \node at (-1,1.2) {$Sp(4,\mathbb R)$};
    \node at (-1,0.4) {$\cup$};
    \node  at (-1,-0.4) {$Sp(2,\mathbb C)$};
    \node at (2,-0.4) {$O(M)$};
    \node at (2,0.4) {$\cup$};
    \node at (2,1.2) {$O(M,\mathbb C)$};
    \end{tikzpicture}
    }.
\end{equation}
For $\mathbb F=\mathbb C$,
the original dual generators form $U(p,M-p)$,
and complex $M_{IJ}$ are antisymmetric 
hence span a $M(M-1)$ dimensional space.
The whole dual group becomes a real form of $O_{2M}$.
For the $p=0$ example, we find
the sea-saw pairs for $\L<0$ and $\L>0$ as 
\begin{equation}
    \parbox{130pt}{
    \begin{tikzpicture}
    \draw [<->] (0,0) -- (1,0.8);
    \draw [<->] (0,0.8) -- (1,0);
    \node at (-1,1.2) {$U(2,2)$};
    \node at (-1,0.4) {$\cup$};
    \node  at (-1,-0.4) {$Sp(4,\mathbb R)$};
    \node at (2,-0.4) {$U(M)$};
    \node at (2,0.4) {$\cup$};
    \node at (2,1.2) {$O(2M)$};
    \end{tikzpicture}
    },
    \qquad \quad 
    \parbox{130pt}{
    \begin{tikzpicture}
    \draw [<->] (0,0) -- (1,0.8);
    \draw [<->] (0,0.8) -- (1,0);
    \node at (-1,1.2) {$U(2,2)$};
    \node at (-1,0.4) {$\cup$};
    \node  at (-1,-0.4) {$Sp(1,1)$};
    \node at (2,-0.4) {$U(M)$};
    \node at (2,0.4) {$\cup$};
    \node at (2,1.2) {$O^*(2M)$};
    \end{tikzpicture}
    }.
\end{equation}

The $M=1$ case provides the conformal particles
having a very special restriction property: it branches into at most two irreps of the isometry subgroup.
For $\mathbb F=\mathbb R$ and $\mathbb C$,
it happens that the conformal particles cover all massless particles.

The $M=2$ case 
provides all ``massive'' particle species in $|\mathbb F|+2$ dimensions, that is, those appearing in 
the dimensional reduction of the most general ``massless particle'' in AdS$_{|\mathbb F|+3}$. 
Here, we used the quotation mark for ``massive'' and ``massless'',
because these cover not only the standard massive and massless particles with compact dual groups, but the analogue ones in tachyonic spectra or BdS with  non-compact dual groups.

To reiterate, for $\L\neq0$,
the restriction 
\eqref{restriction}
to 3d and 4d isometry groups leads, respectively, to the dual pairs,
\be
    (Sp(2,\mathbb R)^2, O(p,2-p)\times O(q,2-q))\,,\qquad 
     (Sp(2,\mathbb C), O(2,\mathbb C))\,,
     \label{3d}
\ee 
and 
\be 
    (Sp(4,\mathbb R), O(p,4-p))\,,\qquad 
    (Sp(1,1), O^*(4))\,.
    \label{4d}
\ee
Therefore, one simply
recovers dual pairs of other type,
which describe
all possible ``massive'' particles in (A)dS$_3$ and (A)dS$_4$, respectively.
For $\L=0$, 
the restriction  provides
an equally good description,
even though it does not correspond
to a reductive dual pair (as the Poincar\'e group is not reductive).

Remark that
the restriction
\eqref{restriction}
to $H_{\L=0}$ reproduces 
the 3d and 4d massive
twistor models
\cite{Fedoruk:2003td,
Bette:2005br,Fedoruk:2005ks,Mezincescu:2013nta,Fedoruk:2014vqa} with the dual groups containing 
respectively $O(2)$ and $U(2)$:
these flat space twistor models
 can be regarded as
the flat limit of 
(A)dS massive twistors described by 
the dual pairs
\eqref{4d} with $p=0$.
The restriction also
reproduces
the 4d continuous spin
twistor model
\cite{Buchbinder:2018soq,Buchbinder:2019iwi,Buchbinder:2019sie},
with the dual group
containing $U(1,1)$\,.
This model can be regarded 
as the flat limit of the AdS continuous spin twistor described by
$(Sp(4,\mathbb R), O(3,1))$.

The particle description provided by  the restriction \eqref{restriction}
has the manifest covariance of the full bulk Lorentz symmetry,
as oppose to the original particle descriptions based on $(B(\O_{\sst (4)}, \mathbb F), B(\eta_{\sst (p,M-p)},\mathbb F))$ having only  manifest boundary Lorentz covariance: for the example of massive particles in AdS$_4$, the original description makes use of real twistor
having only manifest 3d boundary Lorentz covariance, while
the restriction \eqref{restriction}
makes use of complex twistors having manifest covariance of the 
4d bulk Lorentz symmetry.
Note however that the two descriptions are identical: 
they are simply related by a
change of variables.

Let us also mention that
the quantization of twistor models leads to the oscillator representations governed by dual pair correspondences.
These representations are isomorphic to the spinor-helicity representations:
the massless \cite{Dixon:1996wi, Elvang:2013cua, Henn:2014yza, Elvang:2015rqa} and massive 
\cite{Conde:2016izb, Conde:2016vxs, Arkani-Hamed:2017jhn} in flat space,
as well as their (A)dS counterparts
\cite{ Nagaraj:2018nxq, Nagaraj:2019zmk, Nagaraj:2020sji}
and \cite{Basile:2024ydc}.

For $\mathbb F=\mathbb H$,
the situation is different:
we should require the invariance under $E$
\eqref{E quart} too, since it is a part of the Lorentz algebra.
We find that $M_{IJ}$ no more commute with $E$.
If we had included $M_{IJ}$, which are
$M(2M-1)$ dimensional,
together with the original dual generators
$Sp(p,M-p)$, 
the enhanced dual group would had been
a real form of $GL_{2N}$,
which is dual to a real form of $GL_4$:
for instance, $U(2,2)$ containing the AdS$_5$ group  
or $GL(2,\mathbb H)$ containing dS$_5$ group,  instead of six dimensional isometry group.
The 6d isometry group $H_{\L}$ is a subgroup of $O^*(8)$, at the same time including $SU(2,2)$ and $SL(2,\mathbb H)$,
but not $U(2,2)$ and 
$GL(2,\mathbb H)$.
\begin{equation}
    \parbox{155pt}{
    \begin{tikzpicture}
    \draw [<->] (0.5,0.5) -- (-0.5,-0.5);
    \draw [<->] (1,0.1) -- (-1,-0.1);
    \draw [<->] (0.5,-0.5) -- (-0.5,0.5);
    \draw [left hook-latex] (-1.2,-0.6) -- (-1.2, 0.6);
    \draw [left hook-latex] (-2,0.1) -- (-1.7, 0.63); 
    \draw [right hook-latex] (1.2,-0.6) -- (1.2, 0.6); 
    \draw [right hook-latex]
    (1.7, -0.63) -- (2,-0.1);  
    \node at (-1.2,1) {$O^*(8)$};
     \node at (-2.1,-0.2) {$H_{\L<0}$};
    \node  at (-1.2,-1) {$U(2,2)$};
    \node at (1.2,1) {$U(2M)$};
     \node at (2.1,0.2) {$\tilde H_{\L<0}$};
    \node  at (1.2,-1) {$Sp(M)$};
    \end{tikzpicture}
    },
    \qquad \quad 
   \parbox{155pt}{
    \begin{tikzpicture}
    \draw [<->] (0.5,0.5) -- (-0.5,-0.5);
    \draw [<->] (1,0.1) -- (-1,-0.1);
    \draw [<->] (0.5,-0.5) -- (-0.5,0.5);
    \draw [left hook-latex] (-1.2,-0.6) -- (-1.2, 0.6);
    \draw [left hook-latex] (-2,0.1) -- (-1.7, 0.63); 
    \draw [right hook-latex] (1.2,-0.6) -- (1.2, 0.6); 
    \draw [right hook-latex]
    (1.7, -0.63) -- (2,-0.1);  
    \node at (-1.2,1) {$O^*(8)$};
     \node at (-2.1,-0.2) {$H_{\L>0}$};
    \node  at (-1.2,-1) {$GL(2,\mathbb H)$};
    \node at (1.2,1) {$GL(M,\mathbb H)$};
     \node at (2.1,0.2) {$\tilde H_{\L>0}$};
    \node  at (1.2,-1) {$Sp(M)$};
    \end{tikzpicture}
    }.
\end{equation}

Note that the isometry groups $\widetilde{SO}^+(2,5)$
or $\widetilde{SO}^+(1,6)$ of (A)dS$_6$ are not isomorphic
to another `twistor group', which prevents one from finding
a dual pair containing them in the same symplectic group
in general. Indeed, dual pairs involving, say $O(2,5)$
for the sake of definiteness, are of the form,
\begin{equation}
    \big(O(2,5), Sp(2K,\mathbb R)\big)
    \subset Sp\big(14K,\mathbb R\big)\,,
\end{equation}
for some  $K$,
whereas the twistorial setup for $\mathbb F=\mathbb H$
consists of dual pairs
\begin{equation}
    \big(O^*(8), Sp(M)\big) \subset Sp\big(16M,\mathbb R\big)\,,
\end{equation}
so that in order for the first pair to be contained
in the second one, the rank $K$ and $M$ should verify
$14K=16M$ which forbids low values for them.\footnote{A possible
alternative is to allow for another dual pair,
$\big(H',\tilde H'\big) \subset Sp(2K',\mathbb R)$
such that $7K+K'=8M$. For instance, one could be tempted
to consider the pair
$H'=Sp(1)$ and $\tilde H'=O(1) \cong \mathbb Z_2$,
but its direct product with the dual pair
containing $O(2,5)$ for $K=M$ \emph{is not embedded}
in the $O^*(8)$ dual pair.
Let us also point out that
$O(2,5) \times SU(2)$ is the bosonic subgroup of $F(4)$,
which appears as the symmetry group of Romans supergravity 
\cite{Romans:1985tw}. This may suggest that the seesaw mechanism
discussed here also works in $6d$ in the supersymmetric context.}
}

The impossibility of recovering $H_\L$ with $\mathbb F=\mathbb H$ as part of a dual pair
which in turn forms a seesaw pair with $\big(O^*(8), Sp(M)\big)$
may be understood by recalling that the setup of Howe duality
is quite peculiar from the point of view of dualities
in representation theory. In general, a duality in the context
of representation theory designates a situation wherein
a \emph{reducible} representation $V$ of a given Lie group $G$
is decomposed as a direct sum of tensor products of irreps
of $G$ and irreps of its centralizer, say $\mathcal A$,
in a certain subalgebra of the endomorphism algebra of $V$,
such that pairs of irreps appear only once, and each $G$-irrep
appears with a unique irrep of the $\mathcal A$. The dual pair
correspondence, or Howe duality, corresponds to $V$ being
the Fock space of $N$ pairs of (bosonic) oscillators, 
and the relevant subalgebra of its endomorphism algebra
being the \emph{Weyl algebra}. On top of that,
if $G$ is a reductive subgroup of $Sp(2N,\mathbb R)$,
then its centralizer in the Weyl algebra can be characterized
using invariant theory for classical groups: 
since by assumption $G$ is a product of real forms
of the general linear, orthogonal, or symplectic groups
of various ranks, its invariant are generated by products
of the `elementary invariant tensors' which are
the Kronecker delta, a metric of a given signature
and the canonical symplectic form. Since $Sp(2N,\mathbb R)$
is canonically embedded in the Weyl algebra as its subspace
of quadratic operators (in the oscillators), so is $G$,
and by the previous comment, $G$-invariant operators
are necessarily products of quadratic ones, the latter
being built using the elementary invariant tensors
to contract two oscillators. In other words, the centralizer
of $G$ in the Weyl algebra is generated by quadratic invariants,
and therefore characterizing this centralizer boils down
to identifying the centralizer of $G$ in $Sp(2N,\mathbb R)$%
---the problem can be reduced to studying the quadratic subspace
of the Weyl algebra, i.e. the image of the symplectic group
in it. This latter aspect is the specificity of Howe duality,
and what makes it so appealing and efficient in many context.
However, the general problem of decomposing a space under
the joint action of a group $G$ and its centralizer
(in a certain subalgebra of endomorphism),
though still well-posed, generically cannot be reduced
to the analysis of the centralizer of $G$ in another
finite-dimensional Lie group like the symplectic group.

Consequently, one should not expect that the centralizer
of any Lie group $G$ in the Weyl algebra is generated by
quadratic operator only, even when $G$ is embedded
in quadratic operators: the number of oscillators
should be such that it allows the action of the elementary
$G$-invariant tensors, without leaving `too big'
of an invariant subspace. Such conditions were studied 
by R. Howe in his seminal papers \cite{Howe1989i, Howe1989ii}
containing the classification of reductive dual pairs.
The case of $H_\L$ with $\mathbb F=\mathbb H$ is illustrative of this difficulty,
as even if it is contained as a subgroup in $O^*(8)$,
the Fock space relevant for the latter does not allow
for a decomposition appropriate for invariant tensors
of the former (due to the dimensional mismatch mentioned 
at the beginning of the previous paragraph). This explains
why the centralizer of $H_\L$
for $\mathbb F=\mathbb H$ is generated by a subgroup
contained in the same symplectic group as $O^*(8)$. 
In fact, one should expect its centralizer to be generated
by operators of order higher than quadratic, as is suggested
by the existence of a quartic operator which can be used
to define the mass of particles.
Indeed, non-quadratic but quartic
mass constraints
have been considered in 
\cite{Mezincescu:2010yq,Routh:2015ifa,Mezincescu:2015apa},
and 
it was explained  
in \cite{Arvanitakis:2017cpk}
how these massive Mink$_{3,4,6}$ particles
can be obtained from
the massless AdS$_{4,5,7}$ particles
by adding the mass constraint.
Remark also the recent 
6d flat continuous spin twistor model \cite{Buchbinder:2021bgv}
can be understood as
a restriction of 
an exotic AdS$_7$ massless particle  with
the dual pair 
$(O^*(8), Sp(1,1))$ by a quartic mass constraint.

\acknowledgments

We appreciate the collaboration of
Thomas Basile and 
his contribution on this work. 
The work of E. J. and T. O. was supported by the National Research Foundation of Korea (NRF) grant funded by the Korea government (MSIT) (No. 2022R1F1A1074977). 

\appendix

\section{Quaternion-Symplectic structure}
\label{sec:quaternion}

In this section, 
we provide an alternative way to
handle the quaternion-symplectic structure.
Instead of expanding quaternions by
the typical basis $\mathsf{i}$, $\mathsf{j}$, $\mathsf{k}$,
we may use the $2\times2$ complex matrix representation 
of quaternion:
\begin{equation}
	q=\begin{pmatrix}
	q^1{}_1 & q^1{}_2 \\ 
	q^2{}_1 & q^2{}_2
	\end{pmatrix},
	\qquad 
	\bar q=\begin{pmatrix}
	\bar q^1{}_1 & \bar q^1{}_2 \\ 
	\bar q^2{}_1 & \bar q^2{}_2
	\end{pmatrix},
\end{equation}
with $\bar q^i{}_j=(q^j{}_i)^*$ and additional relations,
\begin{equation}
	q^1{}_1 = (q^2{}_2)^*\,,
	\qquad 
	q^1{}_2 = -(q^2{}_1)^*\,.
	\label{quaternion constraint}
\end{equation}
In this complex matrix description, 
the dot product is given by the trace,
\begin{equation}
	p \bullet q = \frac12\,\bar p^i{}_j\,q^j{}_i
	= \frac12\,\bar q^i{}_j\,p^j{}_i\,.
\end{equation}
The relations \eqref{quaternion constraint} imposed on 
$\xi^\a{}_I{}^i{}_j$ and $\pi_{\dot\a\,I}{}^i{}_j$ 
can be considered as second class constraints,
and hence can be imposed after all the computations 
where we use the Poisson bracket (not the Dirac bracket) given by
\begin{equation}
	\{\xi^\alpha{}_I{}^i{}_k,\bar\pi_{\beta\,J}{}^j{}_l\}
	=\eta_{IJ}\,\delta^\alpha_\beta\,\delta^i_l\,\delta^j_k\,,
	\qquad 
	\{\bar\xi^{\dot\alpha}{}_I{}^i{}_k,\pi_{\dot\beta\,J}{}^j{}_l\}
	=\eta_{IJ}\,\delta^{\alpha}_{\beta}\,\delta^i_l\,\delta^j_k\,,
\end{equation}
where $i,j,k,l=1,2$.

\section{AdS generators}
\label{sec:ads dic}

In this appendix, we provide
the explicit component expressions of the 
$\mathfrak{b}(\O_{\sst (4)},\mathbb F)$ 
generators.

\paragraph{Generators of $\mathfrak{sp}(4,\R)$}
    \be
     J_{0'0}=
     \frac{1}{2}
     \begin{pmatrix}
         0 & 1_{\sst(2)} \\
         -1_{\sst(2)} & 0
     \end{pmatrix},\ 
     J_{0'1}=
     \frac{1}{2}
     \begin{pmatrix}
         0 & \t_2 \\
         \t_2 & 0
     \end{pmatrix},\ 
     J_{0'2}=
    -\frac{1}{2}
     \begin{pmatrix}
         0 & \t_1 \\
         \t_1 & 0
     \end{pmatrix},\ 
     J_{0'3}=
     \frac{1}{2}
     \begin{pmatrix}
         1_{\sst(2)} & 0 \\
         0 & -1_{\sst(2)}
     \end{pmatrix},
    \ee
    \be
    J_{01}=
     \frac{1}{2}
     \begin{pmatrix}
         \t_2 & 0 \\
         0 & -\t_2
     \end{pmatrix},\quad
     J_{02}=
     \frac{1}{2}
     \begin{pmatrix}
         -\t_1 & 0 \\
         0 & \t_1
     \end{pmatrix},\quad
     J_{03}=
     -\frac{1}{2}
     \begin{pmatrix}
         0 & 1_{\sst(2)} \\
         1_{\sst(2)} & 0
    \end{pmatrix},\quad
    \ee
    \be
    J_{12}=
     -\frac{1}{2}
     \begin{pmatrix}
         \t_0 & 0 \\
         0 & \t_0
     \end{pmatrix},\quad
     J_{13}=
     -\frac{1}{2}
     \begin{pmatrix}
         0 & \t_2 \\
         -\t_2 & 0
     \end{pmatrix},\quad
    J_{23}=
     \frac{1}{2}
     \begin{pmatrix}
         0 & \t_1 \\
         -\t_1 & 0
     \end{pmatrix},\quad
    \ee
\paragraph{Generators of $\mathfrak{u}(2,2)$}
    \be
    J_{0'4}=
     \frac{\mathsf{i}}{2}
     \begin{pmatrix}
         0 & \t_0 \\
         \t_0 & 0
     \end{pmatrix},\quad
      J_{34}=
     \frac{\mathsf{i}}{2}
     \begin{pmatrix}
         0 & -\t_0 \\
         \t_0 & 0
     \end{pmatrix},
     \quad 
     J_{14}=
     -\frac{\mathsf{i}}{2}
     \begin{pmatrix}
         \t_1 & 0 \\
         0 & \t_1
     \end{pmatrix},
    \ee
    \be
     J_{04}=
     -\frac{\mathsf{i}}{2}
     \begin{pmatrix}
         \t_0 & 0 \\
         0 & -\t_0
     \end{pmatrix},\quad
     J_{24}=-
     \frac{\mathsf{i}}{2}
     \begin{pmatrix}
         \t_2 & 0 \\
         0 &\t_2
     \end{pmatrix},
     \quad
      J_{56}=
     \frac{\mathsf{i}}{2}
     \begin{pmatrix}
         1_{\sst(2)} & 0 \\
         0 & 1_{\sst(2)}
     \end{pmatrix}.
     \ee
     
\paragraph{Generators of $\mathfrak{o}^*(8)$}
   \be
     J_{0'5}=
     \frac{\mathsf{j}}{2}
     \begin{pmatrix}
         0 & \t_0 \\
         \t_0 & 0
     \end{pmatrix},\quad 
J_{35}=
     \frac{\mathsf{j}}{2}
     \begin{pmatrix}
         0 & -\t_0 \\
         \t_0 & 0
     \end{pmatrix},\quad 
      J_{15}=
     -\frac{\mathsf{j}}{2}
     \begin{pmatrix}
         \t_1 & 0 \\
         0 & \t_1
     \end{pmatrix},
    \ee
\be
 J_{05}=
     -\frac{\mathsf{j}}{2}
     \begin{pmatrix}
         \t_0 & 0 \\
         0 & -\t_0
     \end{pmatrix},\quad  
     J_{25}=-
     \frac{\mathsf{j}}{2}
     \begin{pmatrix}
         \t_2 & 0 \\
         0 & \t_2
     \end{pmatrix},\quad 
      J_{46}=
     -\frac{\mathsf{j}}{2}
     \begin{pmatrix}
         1_{\sst(2)} & 0 \\
         0 & 1_{\sst(2)}
     \end{pmatrix}, 
    \ee
    \be
  J_{0'6}=
     \frac{\mathsf{k}}{2}
     \begin{pmatrix}
         0 & \t_0 \\
         \t_0 & 0
     \end{pmatrix},\quad 
     J_{36}=
     \frac{\mathsf{k}}{2}
     \begin{pmatrix}
         0 & -\t_0 \\
         \t_0 & 0
     \end{pmatrix},\quad 
       J_{16}=
     -\frac{\mathsf{k}}{2}
     \begin{pmatrix}
         \t_1 & 0 \\
         0 & \t_1
     \end{pmatrix},   
\ee
\be 
    J_{06}=
     -\frac{\mathsf{k}}{2}
     \begin{pmatrix}
         \t_0 & 0 \\
         0 & -\t_0
     \end{pmatrix},\quad 
    J_{26}=-
     \frac{\mathsf{k}}{2}
     \begin{pmatrix}
         \t_2 & 0 \\
         0 & \t_2
     \end{pmatrix},\quad 
      J_{45}=
     \frac{\mathsf{k}}{2}
     \begin{pmatrix}
         1_{\sst(2)} & 0 \\
         0 & 1_{\sst(2)}
     \end{pmatrix}. 
\ee

\bibliographystyle{JHEP}
\bibliography{biblio}

\end{document}